\documentclass[apj, iop]{emulateapj}
\usepackage{gensymb}
\usepackage{natbib,twoopt}
\makeatletter
\newcommandtwoopt{\citeads}[3][][]{\href{http://adsabs.harvard.edu/abs/#3}%
{\def\hyper@linkstart##1##2{}%
\let\hyper@linkend\@empty\citealp[#1][#2]{#3}}}
\newcommandtwoopt{\citepads}[3][][]{\href{http://adsabs.harvard.edu/abs/#3}%
{\def\hyper@linkstart##1##2{}%
\let\hyper@linkend\@empty\citep[#1][#2]{#3}}}
\newcommandtwoopt{\citetads}[3][][]{\href{http://adsabs.harvard.edu/abs/#3}%
{\def\hyper@linkstart##1##2{}%
\let\hyper@linkend\@empty\citet[#1][#2]{#3}}}
\newcommandtwoopt{\citeyearads}[3][][]%
{\href{http://adsabs.harvard.edu/abs/#3}
{\def\hyper@linkstart##1##2{}%
\let\hyper@linkend\@empty\citeyear[#1][#2]{#3}}}
\makeatother

\begin{document}

   \title{Observationally based models of penumbral microjets}

   \author{S.~Esteban Pozuelo\altaffilmark{1}, J.~de~la~Cruz~Rodr\'{i}guez\altaffilmark{1}, A.~Drews\altaffilmark{2, 3}, L.~Rouppe~van~der~Voort\altaffilmark{2, 3}, G.~B.~Scharmer\altaffilmark{1} and M.~Carlsson\altaffilmark{2, 3}}

   \altaffiltext{1}{Institute for Solar Physics, Dept. of Astronomy, Stockholm University, AlbaNova University Center, 106 91 Stockholm, Sweden, \\ email: sara.esteban@astro.su.se}
   \altaffiltext{2}{Rosseland Centre for Solar Physics, University of Oslo, P.O. Box 1029 Blindern, NO-0315 Oslo, Norway}
	\altaffiltext{3}{Institute of Theoretical Astrophysics, University of Oslo, PO Box 1029 Blindern, 0315 Oslo, Norway}

  \begin{abstract}

We study the polarization signals and physical parameters of penumbral microjets (PMJs) by using high spatial resolution data taken in the \ion{Fe}{1}~630~nm pair, \ion{Ca}{2}~854.2~nm and \ion{Ca}{2}~K lines with the CRISP and CHROMIS instruments at the Swedish 1-m Solar Telescope. We infer their physical parameters, such as physical observables in the photosphere and chromospheric velocity diagnostics, by different methods, including inversions of the observed Stokes profiles with the STiC code. PMJs harbor overall brighter \ion{Ca}{2}~K~line profiles and conspicuous polarization signals in \ion{Ca}{2}~854.2~nm, specifically in circular polarization that often shows multiple lobes mainly due to the shape of Stokes~$I$. They usually overlap photospheric regions with sheared magnetic field configuration, suggesting that magnetic reconnections could play an important role in the origin of PMJs. The discrepancy between their low LOS velocities and the high apparent speeds reported on earlier, as well as the existence of different vertical velocity gradients in the chromosphere, indicate that PMJs might not be entirely related to mass motions. Instead, PMJs could be due to perturbation fronts induced by magnetic reconnections occurring in the deep photosphere that propagate through the chromosphere. This reconnection may be associated with current heating that produces temperature enhancements from the temperature minimum region. Furthermore, enhanced collisions with electrons could also increase the coupling to the local conditions at higher layers during the PMJ phase, giving a possible explanation for the enhanced emission in the overall \ion{Ca}{2}~K profiles emerging from these transients.

\end{abstract}


\section{Introduction}
\label{sec:intro}

Penumbral microjets (PMJs) were first reported by \citetads{2007Sci...318.1594K} from time sequences obtained in the \ion{Ca}{2}~H line with the Solar Optical Telescope (SOT; \citeads{2008SoPh..249..167T}) on board the Hinode satellite \citepads{2007SoPh..243....3K}. They are observed as short-lived, small, elongated transients whose apparent rise speeds are of order 100~km~s$^{-1}$. Their average length, width and lifetime are about 640~km, 210~km and 90~s, respectively \citepads{2017A&A...602A..80D}.  

PMJs stand out from their surroundings since they are 10--20\% brighter in the Hinode \ion{Ca}{2}~H passband. However, their orientation with respect to the penumbral background shows a clear center-to-limb variation. Towards the limb, PMJs pop up at an angle relative to the penumbral filaments and it is easy to detect them, while at disk center transients and filaments are aligned and their detection becomes more difficult. In addition, PMJs are upward directed in the photosphere with elevation angles in the range 20~to~60\degree\, and gradually become more horizontal higher up as they follow magnetic field lines within the flux tubes forming the spot \citepads{2008A&A...488L..33J}.

Another interesting aspect concerns the intensity profiles in the \ion{Ca}{2}~854.2~nm line emerging from PMJs. They show a distinctive mild moustache shape \citepads{1956Obs....76..241S} with emission peaks at about $\pm$0.3~\AA\, and a relatively unperturbed absorption line core, reminiscent of line profiles of Ellerman bombs \citepads{1917ApJ....46..298E}. According to \citetads{2013ApJ...779..143R}, this suggests that the event that triggers PMJs is located around the temperature minimum region.

Since their discovery, magnetic reconnection has been proposed as the driving mechanism of PMJs. In the photosphere, sunspot penumbrae have a complex magnetic field topology, commonly referred to as the spine-intraspine structure \citepads{1993ApJ...418..928L}. This configuration corresponds to extended lanes of relatively weak horizontal magnetic field (intraspines) that are embedded in a background medium where the magnetic field is stronger and more vertical (spines). Hence, PMJs could be the consequence of reconnection between differently inclined magnetic field lines. Specifically, \citetads{2016ApJ...816...92T} suggested that magnetic reconnections take place between spines and opposite polarity fields existing at the lateral sides (\citeads{2013A&A...553A..63S}, \citeads{2013A&A...549L...4R}, \citeads{2013A&A...557A..25T}) and at the outer end of the outflow channels. Magnetohydrodynamic simulations back up the magnetic reconnection scenario (\citeads{2008ApJ...687L.127S}, \citeads{2010ApJ...715L..40M}). 

Alternatively, \citetads{2008ApJ...686.1404R} proposed that PMJs are due to shocks originated by magnetic reconnections occuring between neighboring penumbral filaments, which is consistent with the precursor phase of $\sim$1~minute reported by \citetads{2013ApJ...779..143R}. 

Previous studies seem to include magnetic reconnection as a necessary ingredient to explain the observations. \citetads{2010A&A...524A..20K} and \citetads{2016ApJ...816...92T} reported small-scale downflows and opposite polarity patches in the photosphere both related to PMJs, respectively, while a progressive heating to transition region (TR) temperatures along PMJs was reported by \citetads{2015ApJ...811L..33V}. These findings could be signatures of bi-directional flows produced by the magnetic reconnection process.

However, \citetads{2017ApJ...835L..19S} recently found some bright dots above a sunspot in the TR related to PMJs observed in the chromosphere and proposed that the reconnection process could actually happen at the lower corona-TR, i.e., further up in the solar atmosphere than hitherto thought.

Nowadays, we have sufficient knowledge about the morphology of PMJs to speculate about their origin based on previous studies of intensity data acquired in chromospheric lines and polarization signals from the photosphere. However, as far as we know, spectropolarimetric information from the chromosphere has not been analyzed yet.

Our present study is focused on aspects that have so far remained unexplored. Specifically, we examine for the first time if polarization signals emerging from PMJs leave any distinctive imprint in the chromosphere, their fine structure and physical properties, by inspecting high spatial resolution and high temporal cadence data acquired between the photosphere and the chromosphere with the CRISP and CHROMIS instruments at the Swedish 1-m Solar Telescope. Included in these data are the first spectrally resolved \ion{Ca}{2}~K observations at the highest spatial resolution available.

This paper is organized as follows. The observations and the data reduction process are described in Section~\ref{sec:observation}. We define how PMJs were detected in Section~\ref{sec:detection}. The characterization of the polarization in PMJs, their spatial distribution, and the inference of some of their atmospheric parameters obtained by applying inversions and other methods are outlined in Section~\ref{sec:results}. After that, our results are interpreted, discussed and compared with previous studies in Section~\ref{sec:discussion}. Section~\ref{sec:summary} gives an overview of our work.

\section{Observation and data reduction}
\label{sec:observation}

\begin{figure}[!t]
\centering
\includegraphics[trim = {2.2cm 2cm 2.4cm 3.1cm}, clip, width = 0.48\textwidth]{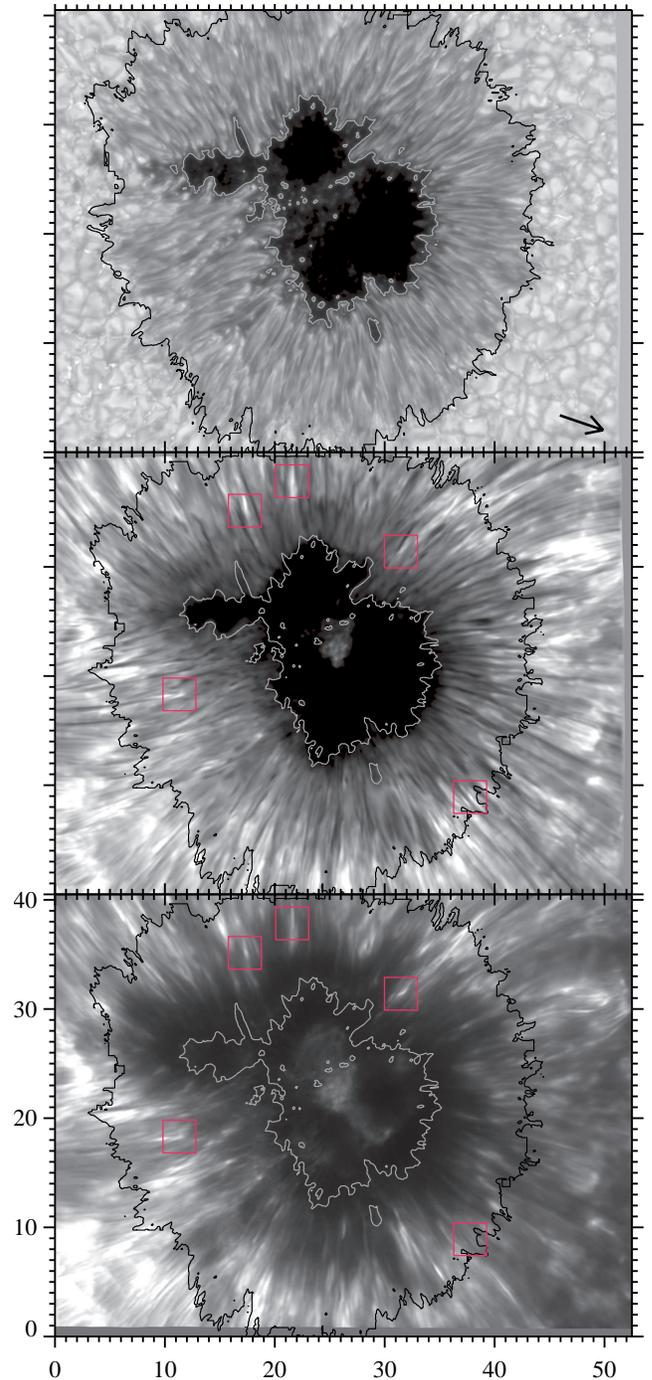} 
\caption{Filtergrams of continuum intensity image of the \ion{Fe}{1}~630.1~nm line (upper), blue-wing of \ion{Ca}{2}~854.2~nm at $-$0.21~\AA\ (middle) and blue wing of \ion{Ca}{2}~K at $-$0.24~\AA\, (lower panel). This scan was taken at 09:55:00~UT. Pink squares show positions of PMJs in the mid and lower panels. Axes are in arcsec. Contours outline the inner and outer boundaries of the penumbra and the arrow points to the solar disk center.}
\label{fig:data}
\end{figure}

The main sunspot in the active region 12585 was observed on September 5$^{th}$, 2016 between 09:48:31 and 10:07:24 UT (upper panel in Figure~\ref{fig:data}) with the CHROMospheric Imaging Spectrometer (CHROMIS) and the CRisp Imaging SpectroPolarimeter \citepads[CRISP;][]{2006A&A...447.1111S, 2008ApJ...689L..69S} at the Swedish 1-m Solar Telescope \citepads[SST;][]{2003SPIE.4853..341S}. The spot had negative polarity and was located at a heliocentric angle of~7.5$^{\degree}$ ($\mu$ = 0.99). 

The CRISP data consist of a multi-wavelength time series of full-Stokes measurements in the photospheric \ion{Fe}{1}~630~nm pair and the chromospheric \ion{Ca}{2}~854.2~nm lines at high spatial resolution and temporal cadence. The \ion{Ca}{2}~854.2~nm line was sampled at 19~line positions between $\pm$100~pm from the line core, plus additional two positions at $\pm$175~pm. As the wings of the \ion{Ca}{2}~854.2~nm line have a photospheric contribution and its core is mainly formed in the chromosphere, non-equidistant sampling steps of 7~pm were used close to core and a wider sampling in the wings. The \ion{Fe}{1}~630~nm pair sampling consists of non-equidistant 16~line positions between~$-$155~and~$+$10~pm from the \ion{Fe}{1}~630.2 nm~line core. Data in the \ion{H}{1}~656.3~nm (H$\alpha$) line were also acquired, but not used in this study. A complete sampling of the three spectral lines required 32~s. The duration of this series is $\sim$20~minutes (35 timesteps). The whole FOV is 58\arcsec\, x 58\arcsec\, with a pixel size of 0\farcs057.

The CHROMIS data are a temporal sequence of high spatial resolution filtergrams in the \ion{Ca}{2}~K~393.4~nm line. 
This line was sampled at 18~line positions in steps of 6~km~s$^{-1}$ between $\pm$50~km~s$^{-1}$ from the line center, plus two extra points at $\pm$100~km~s$^{-1}$ and a continuum point at 400~nm. In addition, data in \ion{H}{1}~486.1~nm (H$\beta$) were also obtained, but not used in the present work. The temporal cadence of the CHROMIS data is $\sim$14~s, which is less than half that of CRISP. The CHROMIS sequence has 91 timesteps. Images have a FOV of 68\arcsec\, x 44\arcsec\, and a pixel size of 0\farcs0375. 

The data reduction was performed individually for each spectral line. The CRISP data were reduced using the CRISPRED pipeline \citepads{2015A&A...573A..40D}. We used a new data reduction pipeline called CHROMISRED \citepads{2018arXiv180403030L} to process the CHROMIS data. All data were processed with the Multi-Object-Multi-Frame-Blind-Deconvolution image restoration technique \citepads[MOMFBD;][]{1994A&AS..107..243L, 2005SoPh..228..191V}. For the CRISP data, the polarimetric calibration was performed for each pixel of the FOV following \citetads{2008A&A...489..429V}. 

During the alignment of both datasets, we took into account the differences in cadence and image scale between the CRISP and CHROMIS datasets. As the temporal cadence of the CHROMIS data is higher, CRISP images were duplicated and systematically destreched to have the same number of images on both datasets and facilitate the subsequent analysis. Finally, the CRISP and CHROMIS datasets were accurately aligned by scaling up the CRISP pixel scale to that of CHROMIS (0\farcs0375) and cross-correlating images acquired at the continuum in the \ion{Ca}{2}~854.2~nm and in the \ion{Ca}{2}~K.

\section{Detection of PMJs}
\label{sec:detection}

\begin{figure}[!t]
\centering
\includegraphics[trim = {0cm 2.5cm 0cm 1cm}, clip, width = 0.47\textwidth]{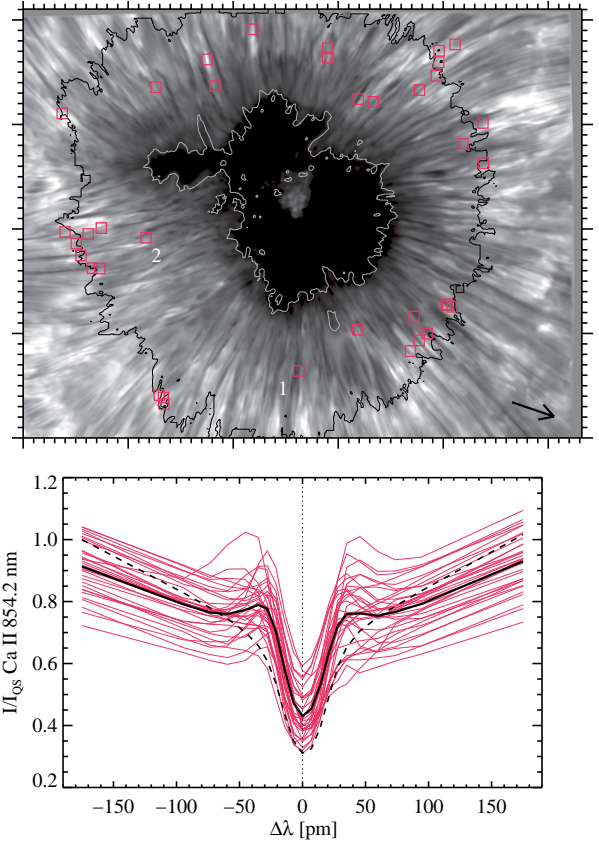} 
\caption{Positions and \ion{Ca}{2} 854.2~nm intensity profiles of the detected PMJs. \textit{Top}: Location of our 37 examples overplotted on a blue-wing filtergram of the \ion{Ca}{2} 854.2~nm line (--0.21\AA). Each major tick represents 10\arcsec. \textit{Bottom}: \ion{Ca}{2} 854.2~nm line profiles emerging from a pixel of each detected PMJ at its maximum brightness and their average (magenta and black solid lines). All intensity profiles are normalized to the average quiet Sun intensity (I$_{QS}$) in the first observed wavelength. The mean quiet Sun profile in the scan taken at 09:55:00~UT is plotted as a reference (black dashed).}
\label{fig:detection}
\end{figure}
 
PMJs are small, bright, jet-like transients in chromospheric penumbrae, whose morphology, life dynamics and distintive \ion{Ca}{2}~854.2~nm line profiles have been previously reported (see Section~\ref{sec:intro}). Considering these aspects, we detected PMJs as elongated brightenings through visual inspection of time series of blue- and red-wing filtergrams in the \ion{Ca}{2}~854.2~nm line (at $\pm$ 0.35~\AA) using the CRIsp SPectral EXplorer (CRISPEX, \citeads{2012ApJ...750...22V}) tool. Besides, we checked that their temporal behavior shows a smooth brightness variation \citepads{2013ApJ...779..143R} to ensure that they represent genuine PMJs.

We found 37~PMJs, five of them are marked in the \ion{Ca}{2} 854.2~nm and \ion{Ca}{2} K maps displayed in Figure~\ref{fig:data}. Although they are clearly seen in the \ion{Ca}{2} 854.2~nm filtergram, some of them are more noticeable in \ion{Ca}{2} K due to a higher contrast between the sharp, bright PMJs and the dark penumbral background.



Regarding their appearance, these PMJs exhibit a wide range of lengths from 725 to 2500~km and widths between 290 and 750~km, with median values of 1450 and 435~km, respectively. They can be observed for 4--28~frames ($\sim$1--6.5~minutes); some of them are recurrent.

Figure~\ref{fig:detection} shows positions and \ion{Ca}{2} 854.2~nm line profiles emerging from a pixel at the maximum brightness of the detected PMJs. They are located on both penumbral sides at different radii, being more visible at the outer boundary. Their intensity profiles in \ion{Ca}{2} 854.2~nm (magenta lines) are well-defined with emission peaks in one or both wings, where the intensity signals are 10--60\% greater than for the quiet Sun (QS). Their average profile (black solid) facilitates the display of this signature, in particular when comparing with the QS profile (black dashed).

\section{Results}
\label{sec:results}

In this section we characterize the polarization in PMJs, the spatial distribution of their \ion{Ca}{2}~854.2~nm Stokes profiles, and some of their physical properties by means of inversions and other techniques. Since all examples have similar characteristics, we illustrate them by showing two cases located in different positions (labeled as 1 and 2 in Figure~\ref{fig:detection}).

We refer to the profile features of the \ion{Ca}{2}~K line as in \citetads{1904ApJ....19...41H} and define the central minimum as K$_{3}$ and the two emission peaks as K$_{2V}$ and K$_{2R}$, where $V$ and $R$ stand for the violet (blue) and red side of the line.

\subsection{Chromospheric polarization signals in PMJs}
\label{sub:pol}

\begin{figure}[!t]
\centering
\includegraphics[trim = {0.75cm 0.9cm 1.2cm 3.0cm}, clip, width = 0.49\textwidth]{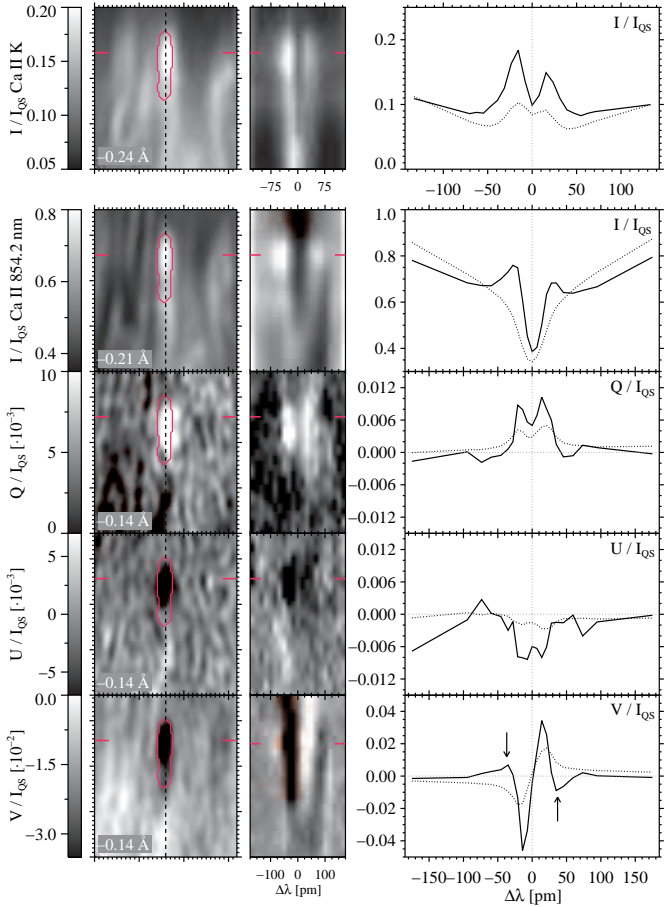} 
\caption{Details of PMJ~1. \textit{Left:} Close-ups of the intensity \ion{Ca}{2}~K at --0.24~\AA, intensity \ion{Ca}{2}~854.2~nm at --0.21~\AA\ filtergrams and blue-wing polarization maps of \ion{Ca}{2}~854.2~nm at --0.14~\AA. Each major tick mark represents 1\arcsec. Contours enclose the extension of the PMJ.  \textit{Middle:} Spectra emerging along the dashed line drawn in the close-ups. \textit{Right:} Stokes profiles observed at the position of the PMJ marked with horizontal magenta dashes on both sides of the previous panels (solid line). For comparison, the average Stokes profiles computed using the pixels surrounding the PMJ are also plotted (dotted line). Arrows point out two \textit{extra} lobes in Stokes~$V$.}
\label{fig:example1}
\end{figure}

\begin{figure}
\centering
\includegraphics[trim = {0.75cm 1.2cm 1.6cm 3.3cm}, clip, width = 0.49\textwidth]{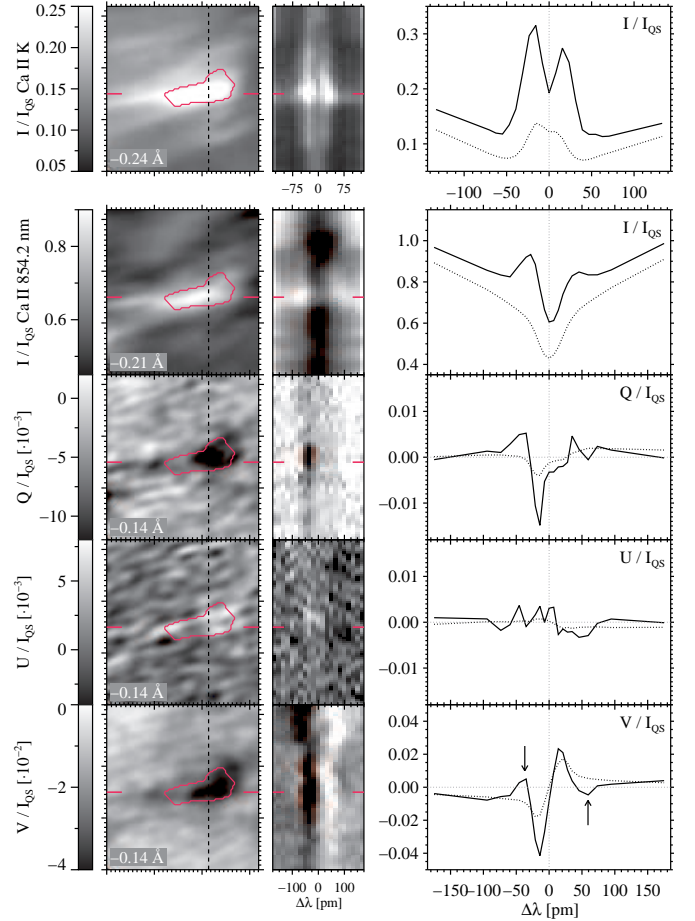} 
\caption{Details of PMJ~2. The layout is the same as in Figure~\ref{fig:example1}.}
\label{fig:example4}
\end{figure}

For the first time, we study the spectropolarimetric information in the \ion{Ca}{2}~854.2~nm line emerging from pixels harboring PMJs. In the case of the \ion{Ca}{2}~K data, we study the intensity profiles.

Figure~\ref{fig:example1} corresponds to PMJ~1, which is located in the outer part of a disk-center penumbral region close to the perpendicular direction to the symmetry line, the line that connects the center of the sunspot to disk center. The PMJ is observed as a bright, elongated feature in the intensity panels. Moreover, it coincides with clearly enhanced signals in polarization that unveil the existence of differently inclined magnetic field lines within the PMJ. 

In the middle column of Figure~\ref{fig:example1}, the intensity spectra of the \ion{Ca}{2}~854.2~nm and \ion{Ca}{2}~K lines along the feature exhibit brightenings in both wings at the position of the PMJ, being greater in the blue wing than in the red one. All polarization spectra reveal enhanced signals in the PMJ, being remarkably irregular in Stokes~$V$ (see further discussion below). In the right column, the corresponding intensity profile of \ion{Ca}{2} 854.2~nm exhibits an expected moustache shape, whose blue emission peak is larger than the red one. Regarding the overall \ion{Ca}{2}~K intensity profile, it is considerably brighter and wider than that on the average emerging from the surroundings. The K$_{2}$~peaks show the same asymmetry as the emission peaks of the \ion{Ca}{2}~854.2~nm line, suggesting that the PMJ leaves similar imprints on both spectral ranges, i.e., between the temperature minimum region and the mid chromosphere (\citeads{2008A&A...480..515C}, \citeads{2009A&A...500.1239R}, \citeads{2018A&A...611A..62B}). Furthermore, the polarization profiles are well above the noise level. The linear polarization signals do not show distinct features, except they are stronger and more well-defined than in their surroundings, in particular as regards Stokes $Q$. The circular polarization shows a four-lobed profile.

Our second example, PMJ~2, is located in the mid limb-side penumbra and is represented in Figure~\ref{fig:example4}. We can essentially notice similar aspects to those outlined in Figure~\ref{fig:example1}, even though its Stokes~$U$ signal is poorer. 

\begin{figure}[!t]
\centering
\includegraphics[trim = {0.6cm 0.8cm 2cm 0cm}, clip, width = 0.48\textwidth]{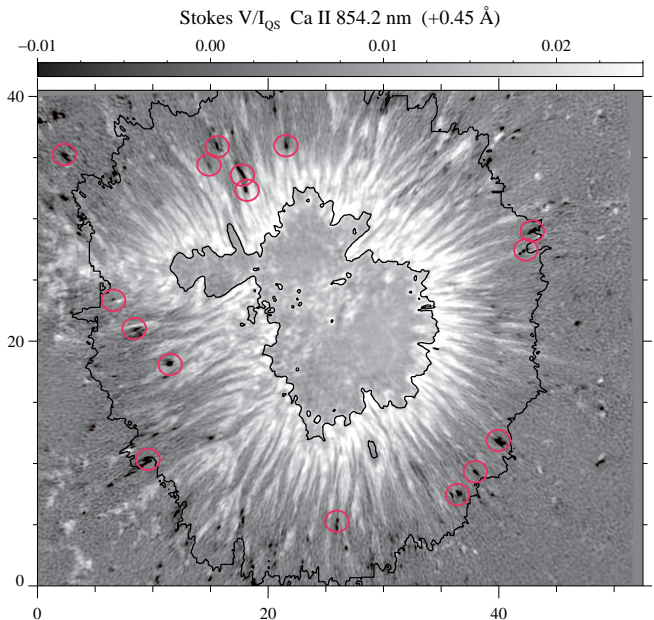} 
\caption{Far red-wing magnetogram of the \ion{Ca}{2}~854.2~nm line. Magenta circles show the position of enhanced signal coinciding with PMJs. The black contours outline the penumbral region. Axes are labeled in arcsec.}
\label{fig:magnetogram}
\end{figure}

Both examples have four-lobed Stokes $V$ profiles in the \ion{Ca}{2}~854.2~nm line. Figure~\ref{fig:magnetogram} shows a far red-wing magnetogram of the \ion{Ca}{2}~854.2~nm line (at $+$0.45~\AA), where conspicuous black patches are observed, mainly in the mid and outer penumbra. Far blue-wing magnetograms also show greater signals, however they look more prominent in the far red-wing by contrast. We find many of these patches to coincide with the location of PMJs (marked by magenta circles), while the few remaining are related to other features. Thus, far-wing magnetograms are a good diagnostic to identify PMJs. Particularly in the case of sunspots close to the disk center, identifying PMJs in far-wing magnetograms is more effective than in intensity diagnostics since due to projection, PMJs in intensity are closely aligned to background photospheric penumbral filaments of comparable intensity level and PMJs identification becomes very challenging. In the far-wing magnetograms, PMJs stand out clearly as opposite polarity features for sunspots at any location on the disk.

We interpret the shape of the \ion{Ca}{2}~854.2~nm Stokes~$V$ profiles emerging from PMJs by using the weak-field approximation (WFA), which assumes that the magnetic field strength is constant with depth and weak enough to produce a Zeeman splitting much smaller than the Doppler width throughout the formation region of a spectral line \citepads{2004ASSL..307.....L}. In previous studies, the WFA has been successfully applied in chromospheric data (as in \citeads{2012MNRAS.419..153M}), particularly in \ion{Ca}{2}~854.2~nm Stokes~$V$ observations (\citeads{2013A&A...556A.115D}, \citeads{2017A&A...599A.133A}, and others) because it is a relatively broad line and its Land\'{e} factor is relatively low ($\tilde{g}$ = 1.10). 

According to the WFA, the observed Stokes~$V$ profile of a spectral line can be expressed as

\begin{equation} \label{eq:1}
V(\lambda)\, = \, -C B_{\parallel} \frac{\partial I(\lambda)}{\partial \lambda}
\end{equation}

\noindent where $C$ is a constant dependent on the laboratory wavelength and the effective Land\'{e} factor, $B_{\parallel} = B~cos ~\gamma$ is the longitudinal component of the magnetic field that coincides with the magnetic field along the line-of-sight (B$_{LOS}$), $B$ is the field strength, $\gamma$ is the inclination of the field with respect to the line-of-sight (LOS) and $I$ is the observed intensity profile. Thus, the shape of Stokes~$V$ is given by the partial derivative of Stokes~$I$ and the role of $B_{\parallel}$ (times $-C$) is to play as a scale factor. 

\begin{figure}[!t]
\centering
\includegraphics[trim = {1.3cm 1cm 1.3cm 3cm}, clip, width = 0.45\textwidth]{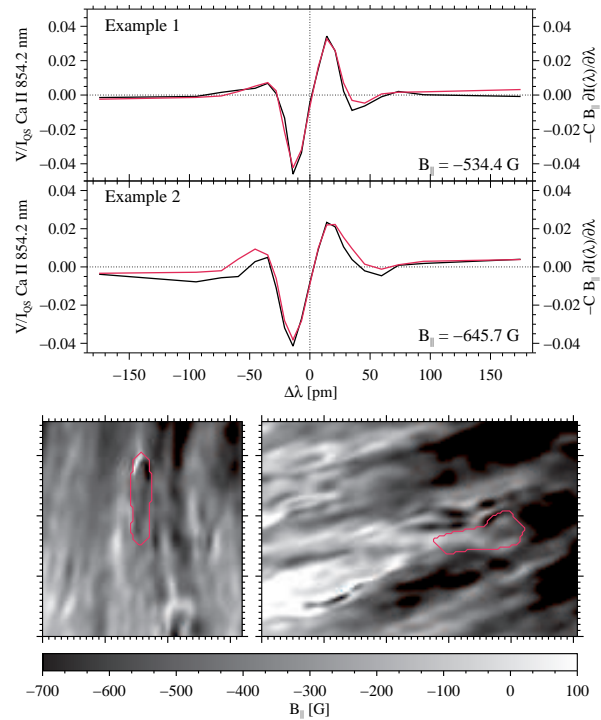} 
\caption{Results from the WFA on PMJs~1 and~2. \textit{Top:} Comparison of the observed Stokes~$V$ and $\partial I(\lambda) / \partial\lambda$ in \ion{Ca}{2}~854.2~nm (black and magenta lines) of the pixels shown in Figures~\ref{fig:example1} and~\ref{fig:example4}. The $B_{\parallel}$ value inferred from Equation~\ref{eq:2} is indicated in the lower right corner of each panel. \textit{Bottom:} Maps of the $B_{\parallel}$ retrieved for PMJs~1 and~2 (left and right) by applying the WFA on wavelengths corresponding to the \textit{outer} lobes of the Stokes~$V$ profile. Contours enclose each PMJ. Each major tick mark represents 1\arcsec.}
\label{fig:wfa}
\end{figure}

An estimated value of B$_{\parallel}$ can be inferred by applying a linear least-squares fit to solve Equation~\ref{eq:1}, which yields

\begin{equation} \label{eq:2}
B_{\parallel}\, = \, \frac{\sum_{i}\frac{\partial I(\lambda_{i})}{\partial\lambda_{i}} V(\lambda_{i})}{-C \, \sum_{i} \left(\frac{\partial I(\lambda_{i})}{\partial\lambda_{i}}\right)^{2}} \; \; .
\end{equation}

In Figure~\ref{fig:wfa}, we compare the Stokes~$V$ profiles and the partial derivative of the Stokes~$I$ profiles displayed in Figures~\ref{fig:example1} and \ref{fig:example4}. The partial derivative of Stokes~$I$ is scaled by using the $B_{\parallel}$ obtained by Equation~\ref{eq:2}, which point to negative magnetic fields (see the lower right corner of each panel). The derivative of the Stokes~$I$ mimics almost perfectly the Stokes~$V$ profiles in PMJ~1, however, the \textit{outer} lobes in PMJ~2 exhibits some discrepancies, which could reveal gradients of B$_{\parallel}$ in the upper photosphere/low chromosphere. This is supported by the B$_{||}$ values inferred for PMJs~1 and~2 by applying the WFA on the wavelengths corresponding to the \textit{outer} lobes of Stokes~$V$. In the lower panels panels of Figure~\ref{fig:wfa}, the B$_{\parallel}$ values obtained for PMJ~1 are barely distinguishable from those inferred for its surroundings. On the other hand, PMJ~2 shows more noticeable B$_{\parallel}$ values, whose difference with respect to their surroundings seems to mark out the position of the feature.

Other examples show similar behavior between the Stokes~$V$ and the partial derivative of Stokes~$I$. Thus, we conclude that the \textit{outer} lobes of Stokes~$V$ are mainly given by the emission peaks present in Stokes~$I$ and not by the coexistence of multiple magnetic components in the same resolution element. 

\subsection{Spatial variation of Stokes profiles in PMJs}
\label{sub:spatialdistribution}

The fine structure of PMJs spectra is still unknown and, consequently, we do not know if there is any noticeable change within PMJs that could shed more light on them. Here, we report on how the \ion{Ca}{2}~854.2~nm and \ion{Ca}{2}~K intensity profiles change along the length of PMJs during their maximum brightness. 

\begin{figure}[!t]
\centering
\includegraphics[trim = {6.5cm 3.5cm 5cm 7.5cm}, clip, width = 0.45\textwidth]{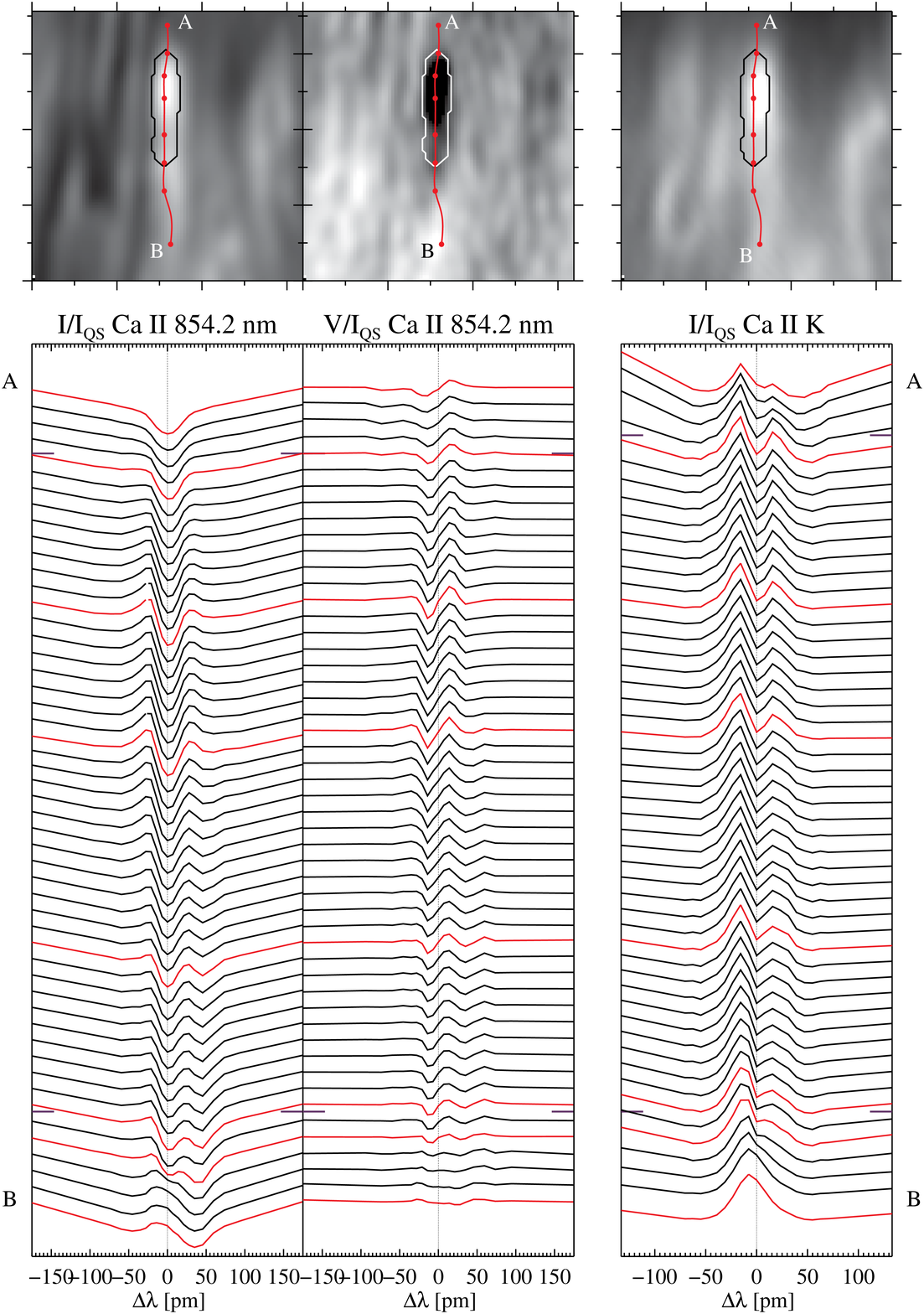} 
\caption{Spatial variation of the Stokes profiles emerging along PMJ~1, corresponding to Type~2 (Table~\ref{tab:types}). \textit{Top:} Close-ups of blue-wing Stokes~$I$ and~$V$ maps in \ion{Ca}{2}~854.2~nm and Stokes~$I$ map in \ion{Ca}{2}~K. Same layout and saturation as in the left column of Figure~\ref{fig:example1}. Contours outline the PMJ. \textit{Bottom:} The observed Stokes profiles of pixels along the cut drawn in the upper close-ups. Horizontal dashes delimit profiles within contours. Vertical lines indicate the rest position. The profiles are arbitrarily shifted in the vertical position.}
\label{fig:spatialexample1}
\end{figure}

\begin{figure}[!t]
\centering
\includegraphics[trim = {8.3cm 4.2cm 8cm 7.7cm}, clip, width = 0.52\textwidth]{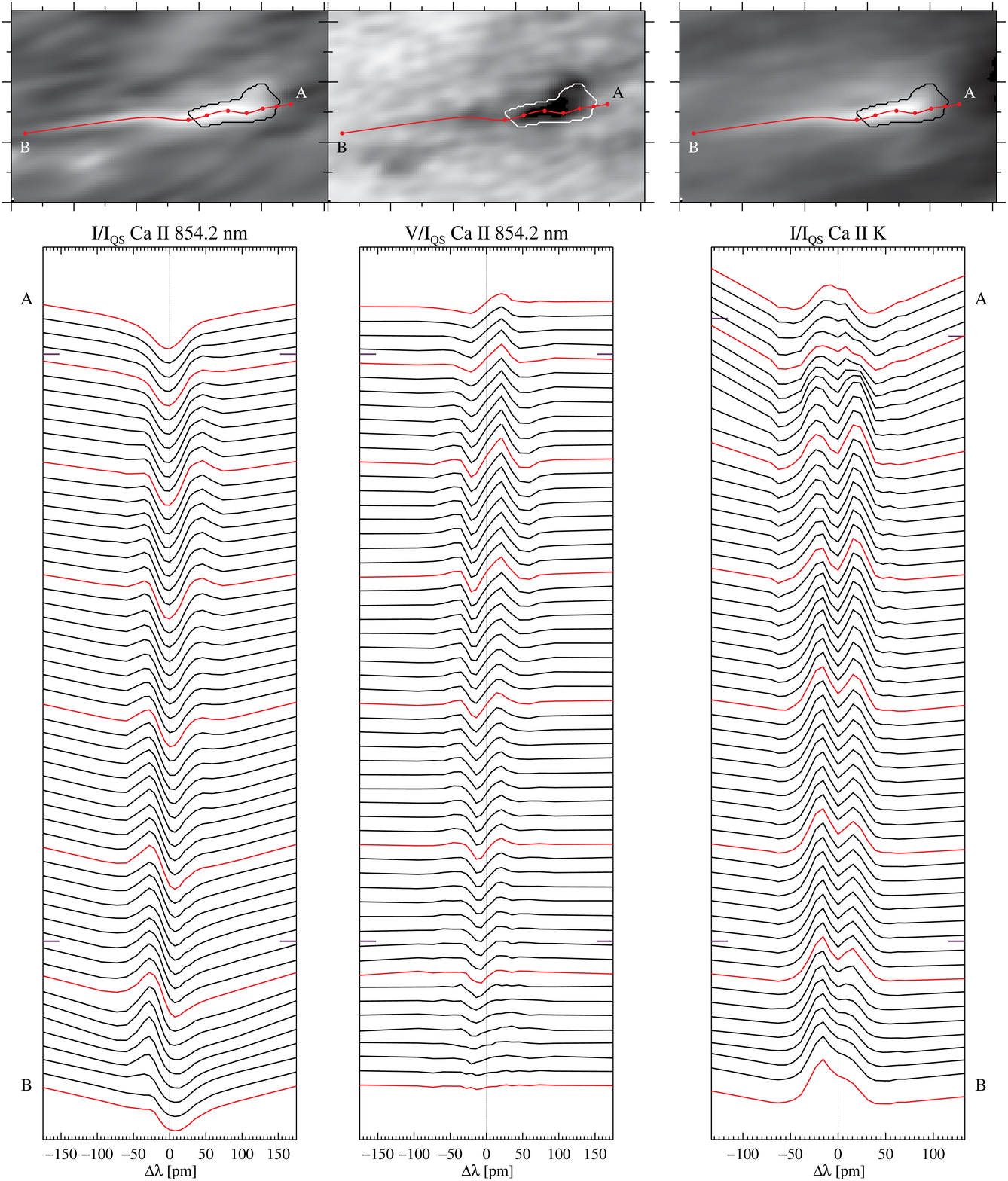} 
\caption{Spatial variation of the Stokes profiles emerging along PMJ~2 showing asymmetry variations in the blue and red wings, corresponding to Type~5 (Table~\ref{tab:types}). Same layout as in Figure~\ref{fig:spatialexample1}.}
\label{fig:spatialexample4}
\end{figure}

\begin{figure}[!t]
\centering
\includegraphics[trim = {0cm 0.2cm 1.6cm 2.5cm}, clip, width = 0.46\textwidth]{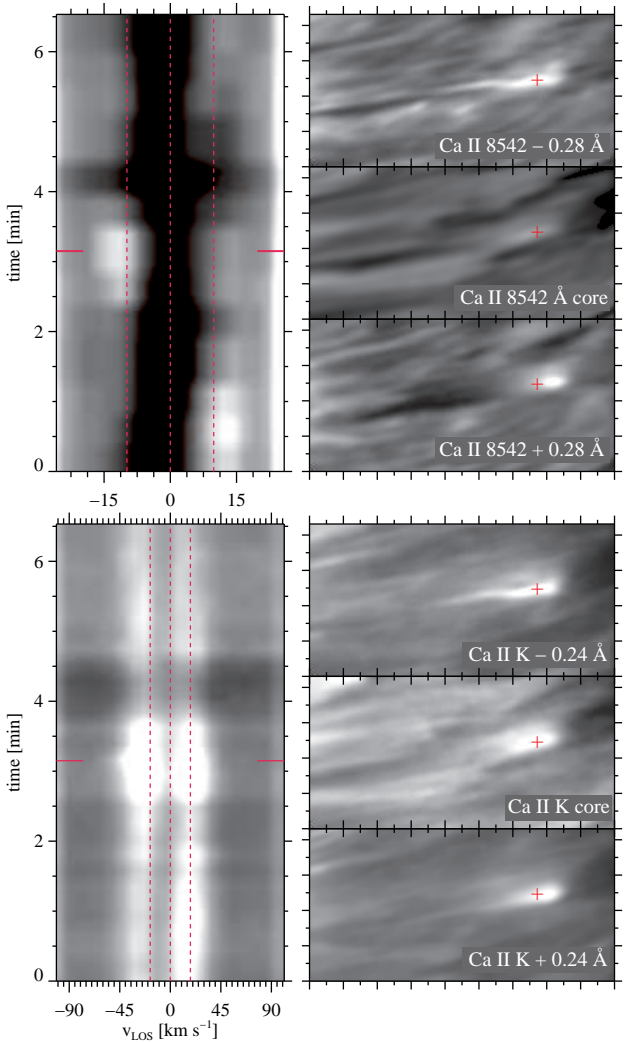} 
\caption{Spectral-temporal behavior of PMJ~2. Spectra-time slices in \ion{Ca}{2}~854.2~nm and \ion{Ca}{2}~K emerging from the pixel described in Figure~\ref{fig:example4}, which is marked with a red plus symbol in the right column. Time goes from bottom to top with a sampling of 14~s. The $x$-axis is expressed as Dopplershift from the line core. Vertical lines and horizontal dashes show the wavelength and temporal positions of the maps displayed on the right side. Each major tick mark in the right column represents 1\arcsec.}
\label{fig:example4time}
\end{figure}

Figure~\ref{fig:spatialexample1} displays all the Stokes profiles emerging along the PMJ~1, but only every fifth pixel outside the contour is shown. The Stokes~$I$ profiles of both lines are not randomly distributed within the PMJ. There is a smooth transition from regular profiles outside this feature to PMJ-like ones along it. The upper Stokes profiles correspond to regular penumbral surrounding pixels, with regular \ion{Ca}{2}~854.2~nm Stokes~$I$ and~$V$ profiles. Although the K$_{2}$ peaks appear enhanced, their signals are actually low and not worthy of mention. After that, \ion{Ca}{2}~854.2~nm intensity profiles develop simultaneously emission peaks on both wings that vary along the PMJ, being more intense in the middle of it. Consequently, Stokes~$V$ signals show four lobes, mostly modulated by the derivative of Stokes~$I$. The K$_{2}$ peaks are also remarkable along the cut, usually displaying amplitude asymmetries similar to those of the \ion{Ca}{2}~854.2~nm emission peaks. Lastly, all signals smoothly decrease as moving far from the center of the PMJ. Unfortunately, the spectral information at the outer end (the one located furthest from the umbra) appears distorted, possibly due to another feature pursuing the PMJ, so we cannot detect how the profiles change to $normal$ ones again.

\begin{table}[t]
\centering
\caption{Spatial variations of the \ion{Ca}{2}~854.2~nm intensity profile along PMJs. The fourth column stands for asymmetry variations along the transients. The total number of PMJs is 37.}
\begin{tabular}{p{0.5cm}cccc}
\textbf{Type}& \textbf{Blue} & \textbf{Red} & \textbf{Asym.} & \textbf{Occurrences}\\
\textbf{} & \textbf{peak} & \textbf{peak} & \textbf{variation} & \textbf{(percentage)}\\
\hline
\hline
1 & Yes & No & No & 7 (18.9\%)\\ 
2 & Yes & Yes (weak) & No & 11 (29.8\%)\\ 
3 & No & Yes & No & 1 (2.7\%)\\  
4 & Yes (weak) & Yes & No & 15 (40.5\%)\\ 
5 & Yes & Yes & Yes & 3 (8.1\%)\\ 
\hline \\
\end{tabular}
\label{tab:types}
\end{table}

We distinguish between five spatial distributions depending on the variation of the \ion{Ca}{2}~854.2~nm intensity profiles along the PMJs, as Table~\ref{tab:types} lists. According to this, PMJ~1 corresponds to Type~2. As in \citetads{2017A&A...602A..80D}, PMJs tend to exhibit a blue emission peak in \ion{Ca}{2}~854.2~nm intensity profiles, which emerges alone or together with the red one, as they usually show Types 1, 2, and 4. The detection of unambiguous Type~5 profiles is not easy by a visual data inspection, so we could retrieve just a few cases, however, its longitudinal asymmetry variation is worthy of study as it is depicted below in PMJ~2.

Figure~\ref{fig:spatialexample4} reveals \ion{Ca}{2}~854.2~nm Stokes~$I$ profiles with emission peaks in the red wing at the inner end of the PMJ (close to point~A). Moving away from the umbra, this red emission peak diminishes while another peak in the blue wing increases, showing similar amplitudes at the center of the PMJ. Then, the red brightening vanishes while the blue one grows. Unfortunately, this PMJ is also affected by another pursuing feature and we cannot observe the transition to a normal intensity profile. The circular polarization shows similar characteristics as before. Regarding the \ion{Ca}{2}~K line, both K$_{2}$ peaks are visible along the longitudinal cut so their amplitudes are not always correlated to those of \ion{Ca}{2}~854.2~nm, as opposed to PMJ~1.

To examine this discrepancy, Figure~\ref{fig:example4time} displays the spectral-temporal behavior of PMJ~2 in the \ion{Ca}{2}~854.2~nm and \ion{Ca}{2}~K lines. The blue wing of \ion{Ca}{2}~854.2~nm is bright while the red wing is covered up by a diagonal dark lane produced by the presence of short dynamic fibrils \citepads{2013ApJ...776...56R} in the line core and red wing. Consequently, only the inner end of the PMJ is discerned since the superpenumbral fibril hides the rest of it at those wavelength positions. Meanwhile, we observe the whole PMJ in all \ion{Ca}{2}~K line filtergrams, except in its core where it is shorter.

\begin{figure}[!t]
\includegraphics[trim = {0.5cm 3.5cm 1.4cm 1.7cm}, clip, width = 0.485\textwidth]{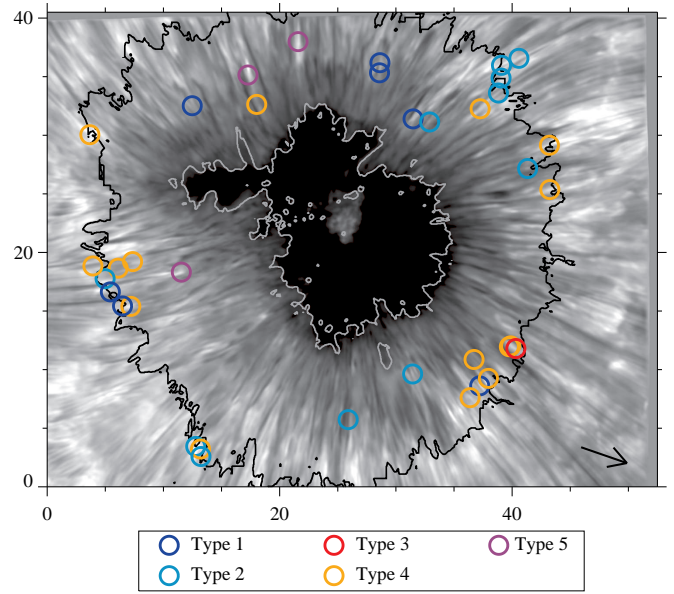} 
\caption{Location of the different spatial variations of intensity profiles along PMJs overplotted on the \ion{Ca}{2}~854.2~nm filtergram shown in Figure~\ref{fig:data}. Different colors distinguish each type (see legend). The black arrow points to the disk center. The black and grey contours outline the penumbral borders. Axes are labeled in arcsec.}
\label{fig:spatialdistsunspot}
\end{figure}

\begin{figure*}[!t]
\centering
\includegraphics[trim = {1.6cm 1.6cm 19cm 0.9cm}, clip, width = 0.86\textwidth]{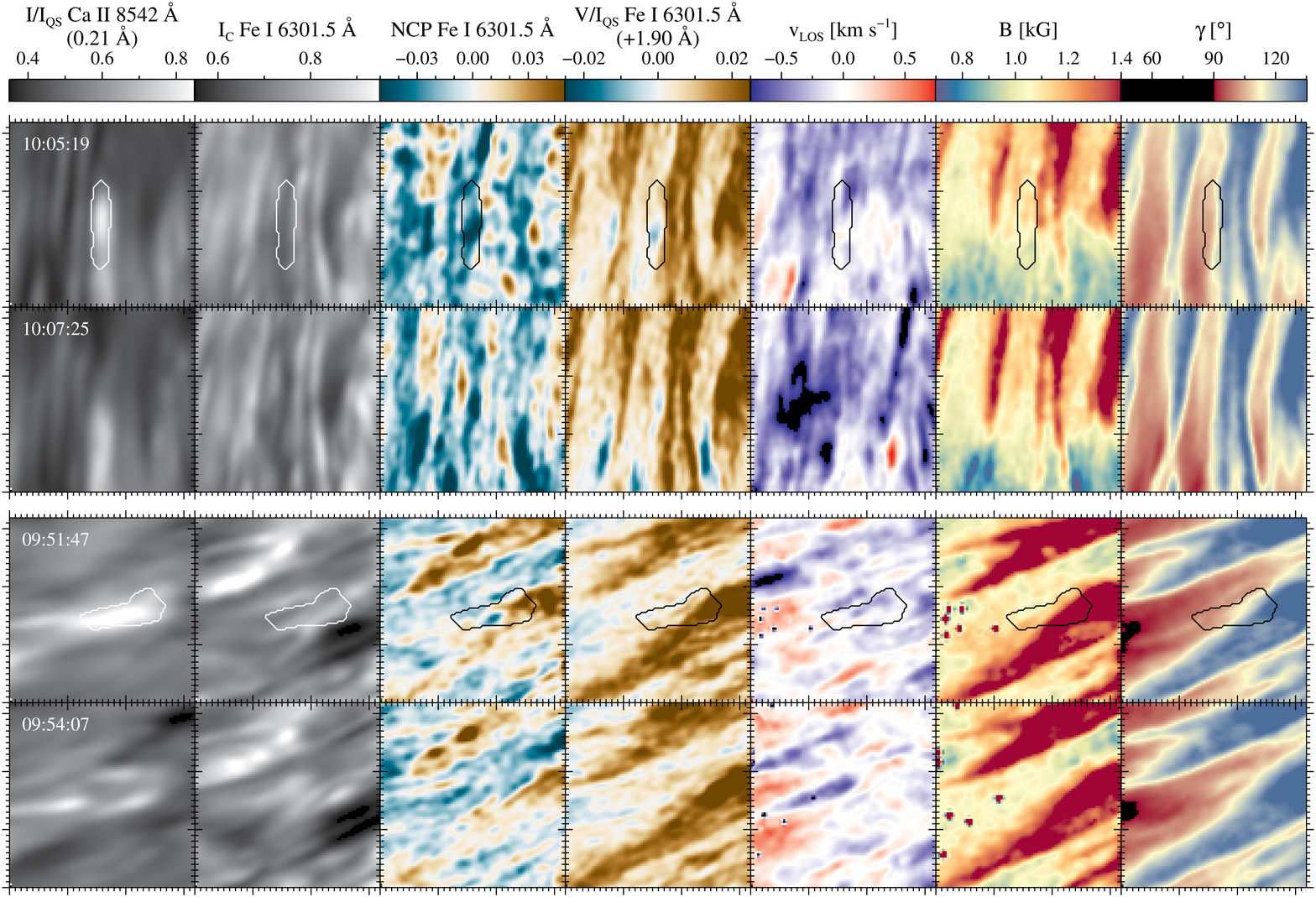} 
\caption{Photospheric diagnostics in the region of PMJ~1 (two top rows) and~2 (two bottom rows). \textit{From left to right}: blue wing intensity filtergram in the \ion{Ca}{2}~854.2~nm line, continuum intensity filtergrams in the \ion{Fe}{1}~630.15~nm~line, NCP maps, red-wing magnetograms of the \ion{Fe}{1}~630.15~nm line, and LOS velocity, magnetic field strength and inclination retrieved from the ME inversions. Contours outline each PMJ case. Each major tick mark represents 1\arcsec.}
\label{fig:photo_pmjs}
\end{figure*}

Therefore, regions located close to the edges of the PMJs reveal a mixture of a PMJ component and another originating from the background during a mild and fast transition. Intensity profiles show distinguishable characteristics that smoothly change and reveal fine changes in the atmospheric parameters within PMJs. However, these signals are easily affected by neighboring features and show distorted shapes. 

We also examined the position of the spatial distributions types throughout the spot, which are indicated by different colored circles in Figure~\ref{fig:spatialdistsunspot}. These distributions appear regardless the penumbral side, so they are not influenced by any projection effect. We also find that Type~4 profiles, with a strong red peak and a weak blue peak, tend to appear close to the outer penumbral boundary.

\subsection{Scenario in the photosphere}
\label{sub:scenario_photosphere}

We analyze the photosphere below PMJs by computing velocity and magnetic field diagnostics. This is motivated by some authors having pointed out the intrincate photospheric magnetic configuration of the penumbra as a possible trigger for PMJs in the chromosphere.

To infer the photospheric LOS velocity and magnetic field below the PMJs, we performed inversions of the Stokes profiles observed in the \ion{Fe}{1} 630~nm pair under the Milne-Eddington (ME) assumption. As the heliocentric angle is small, vertical motions can almost be retrieved from LOS velocity maps. Each Dopplergram was calibrated using as reference the average LOS velocity in the umbra \citepads{1977ApJ...213..900B}. 

In addition, we inspected net circular polarization (NCP) maps of the \ion{Fe}{1}~630.1~nm, which can be used as indicators of gradients in velocity and magnetic fields along our LOS in the photosphere\footnote{Although their presence is not necessary to produce NCP signals, gradients in the magnetic field along the LOS can alter their amount \citepads{1978A&A....64...67A}.} \citepads{1975A&A....41..183I} and reads $NCP \, = \, \int_{\lambda_{1}}^{\lambda_{2}} \, V(\lambda) \, d\lambda$. We also examined far red-wing Stokes~$V$ signals in search of any characteristic signal related to PMJs \citepads{2010A&A...524A..20K}.

Figure~\ref{fig:photo_pmjs} displays the photospheric diagnostics in regions harboring PMJs~1 and~2 in two non-consecutive snapshots: one when the PMJ has its maximum brightness and another only with the pursuing feature. The corresponding blue-wing \ion{Ca}{2}~854.2~nm filtergrams are also shown to indicate the location of the PMJ. 

PMJ~1 appears above the lateral edge of a penumbral filament, specifically in the spine-intraspine transition, where the field strength is about 1.1~kG. Downflows due to magnetic reconnection in the photosphere are expected to be weak \citepads{2010A&A...524A..20K} and, therefore, hard to identify. Within the PMJ~1 contour, we find a patch with negative NCP and red wing Stokes~$V$ values that is almost at rest, coinciding with a field strength gradient. This patch remains approximately at the same place as the PMJ evolves even when only the pursuing feature is visible, suggesting that they are related to other photospheric events, not to the PMJ.

PMJ~2 overlaps a penumbral filament and the lateral edge of a nearby one, also concurring with a spine-intraspine region. Patches with different LOS velocities and NCP values appear with the same orientation as the penumbral filaments within the PMJ. They do not evolve as the PMJ does and can be found in other places as well. Red-wing Stokes~$V$ maps do not show any particular signal within the feature.

Interestingly, PMJs mostly occur where there are strong horizontal gradients in the magnetic field inclination, either between filaments or in the sunspot outer boundary. Intensity filtergrams reveal that PMJs are found above different locations with respect to penumbral filaments, such as on their lateral or outer sides, close to penumbral grains or even overlapping them. Furthermore, many PMJs appear in the sunspot outer boundary, where magnetic field mostly sinks to deeper layers \citepads[e.g.,][]{1997Natur.389...47W} and the Evershed flow partially continues as the moat flow \citepads{1969SoPh....9..347S}. 

Regardless the location of the PMJs over the spot, we can not unambiguously relate them to clear enhanced NCP signals and/or downflows. Subsonic oscillations \citepads{1962ApJ...135..474L} may hide the latter, however, seeing conditions were not stable enough to remove them and filtering produce artifacts.

\subsection{\ion{Ca}{2}~K line diagnostics}
\label{sec:veldiagCaK}

Each profile feature of the \ion{Ca}{2}~K line is sensitive to a different height in the solar atmosphere. The K$_{2}$ and K$_{3}$ peaks are formed in the chromosphere, roughly at 1.0--1.3~Mm and $\sim$2~Mm in the QS, and they provide a powerful tool for LOS velocity diagnosis at these heights \citepads{2018A&A...611A..62B}. 

\begin{figure}[!t]
\centering
\includegraphics[trim = {2.5cm 2cm 2.2cm 2.4cm}, clip, width = 0.375\textwidth]{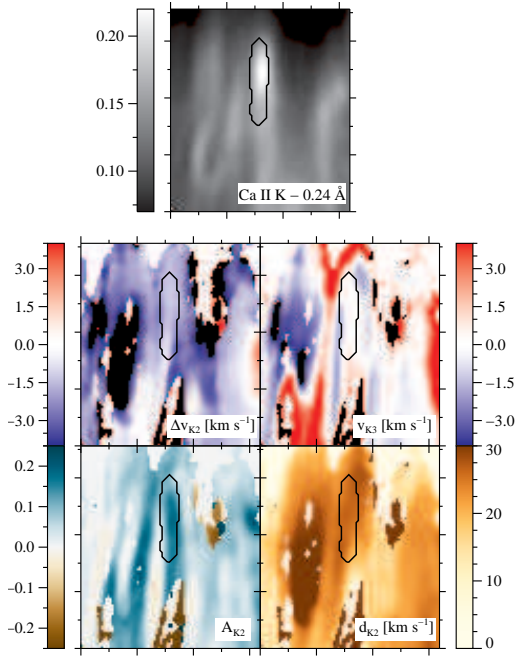} 
\caption{Velocity diagnostics derived from the \ion{Ca}{2}~K peaks for PMJ~1. \textit{Top:} Blue-wing \ion{Ca}{2}~K filtergram. \textit{Bottom:} average Doppler shift of the K$_{2}$~peaks ($\Delta$v$_{K2}$), Doppler velocity of the K$_{3}$ peak (v$_{K3}$), K$_{2}$ peak asymmetry (A$_{K2}$) and their separation (d$_{K2}$). Contours enclose the PMJ. Each major tick mark represents 1\arcsec.}
\label{fig:obscak_pmj1}
\end{figure}

\begin{figure}[!t]
\centering
\includegraphics[trim = {1.5cm 1.2cm 1.2cm 1.2cm}, clip, width = 0.475\textwidth]{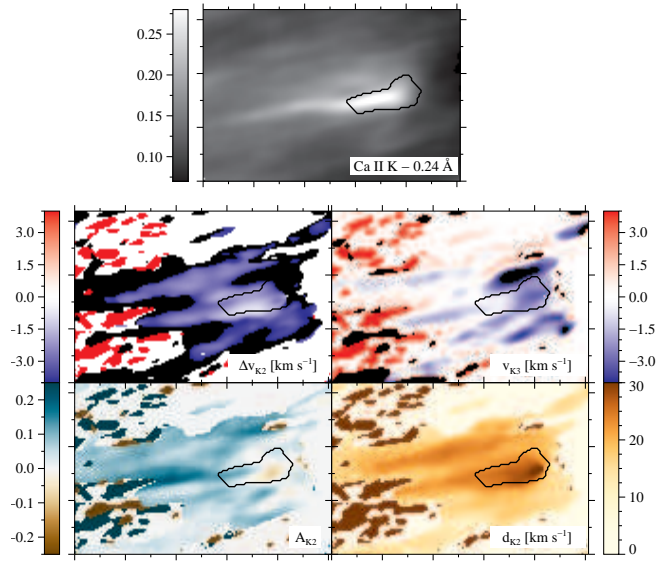} 
\caption{Velocity diagnostics derived from the \ion{Ca}{2}~K peaks for PMJ~2. The layout is the same as in Figure~\ref{fig:obscak_pmj1}.}
\label{fig:obscak_pmj4}
\end{figure}

We applied an automated code to identify the \ion{Ca}{2}~K profile features of each PMJ at their maximum brightness. Following the expressions in \citetads{2018A&A...611A..62B}, we computed the K$_{2}$ peak separation ($d_{K2}$), the asymmetry of the K$_{2}$ peaks ($A_{K2}$), the average Doppler shift of the K$_{2}$ peaks ($\Delta v_{K2}$) and the Doppler velocity of the K$_{3}$ peak ($v_{K3}$) in regions containing each PMJ. We defined the LOS velocity sign criteria, so that negative and positive LOS velocities stand for upflows and downflows, respectively. Figures~\ref{fig:obscak_pmj1} and~\ref{fig:obscak_pmj4} display the results given for PMJs~1 and~2. The identification of the profile features proved satisfactory, although there are some bad fits.

In Figures~\ref{fig:obscak_pmj1} and~\ref{fig:obscak_pmj4}, the average Doppler shift of the K$_{2}$ peaks is blueshifted in almost all the region, even in the PMJs which exhibits upflows of order -1.5~km~s$^{-1}$. However, the Doppler shift of the K$_{3}$~peak is overall more redshifted, reaching values of $\pm$0.6~km~s$^{-1}$ in PMJ~1. The inner end of PMJ~2 is notably more blueshifted to about --2~km~s$^{-1}$. Therefore, there is a velocity decrease between the K$_{2}$ and K$_{3}$ peak formation heights. This velocity gradient is also present in the surroundings of the PMJs. The inner end of PMJ~2 shows an increase in upflow velocity with height.

Figures~\ref{fig:obscak_pmj1} and~\ref{fig:obscak_pmj4} show bright features as lanes with more positive K$_{2}$~peak asymmetries. Among them are the PMJs though there are some differences, as in the inner end of PMJ~2 whose K$_{2}$ peak asymmetries are negative since their K$_{2R}$~peaks are stronger than the K$_{2V}$ ones (see Figures~\ref{fig:spatialexample1} and~\ref{fig:spatialexample4}). Finally, the K$_{2}$~peak separations are greater than 25~km~s$^{-1}$ in both PMJs. This quantity is sensitive to vertical velocity gradients in the upper-mid chromosphere and to temperature enhancements in the lower chromosphere. Thus, the inferred K$_{2}$~peak separations within these PMJs indicate that these effects are present. While the former is usually also found in their surroundings, the latter is distinctive in PMJs as they pop up as sudden brightenings in the lower chromosphere. 

After examining all cases, we observe that flows within PMJs are more blueshifted (or redshifted) with height, if the K$_{2R}$~peak is greater than the K$_{2V}$~one (or viceversa). Thus, different types of \ion{Ca}{2}~K~profiles within a given PMJ correspond to different variations of the vertical velocities between the mid and upper chromosphere, as these are correlated with the K$_{2}$~peak asymmetries. 

\begin{figure}[!t]
\centering
\includegraphics[trim = {1cm 0.4cm 6.5cm 0.5cm}, clip, width = 0.47\textwidth]{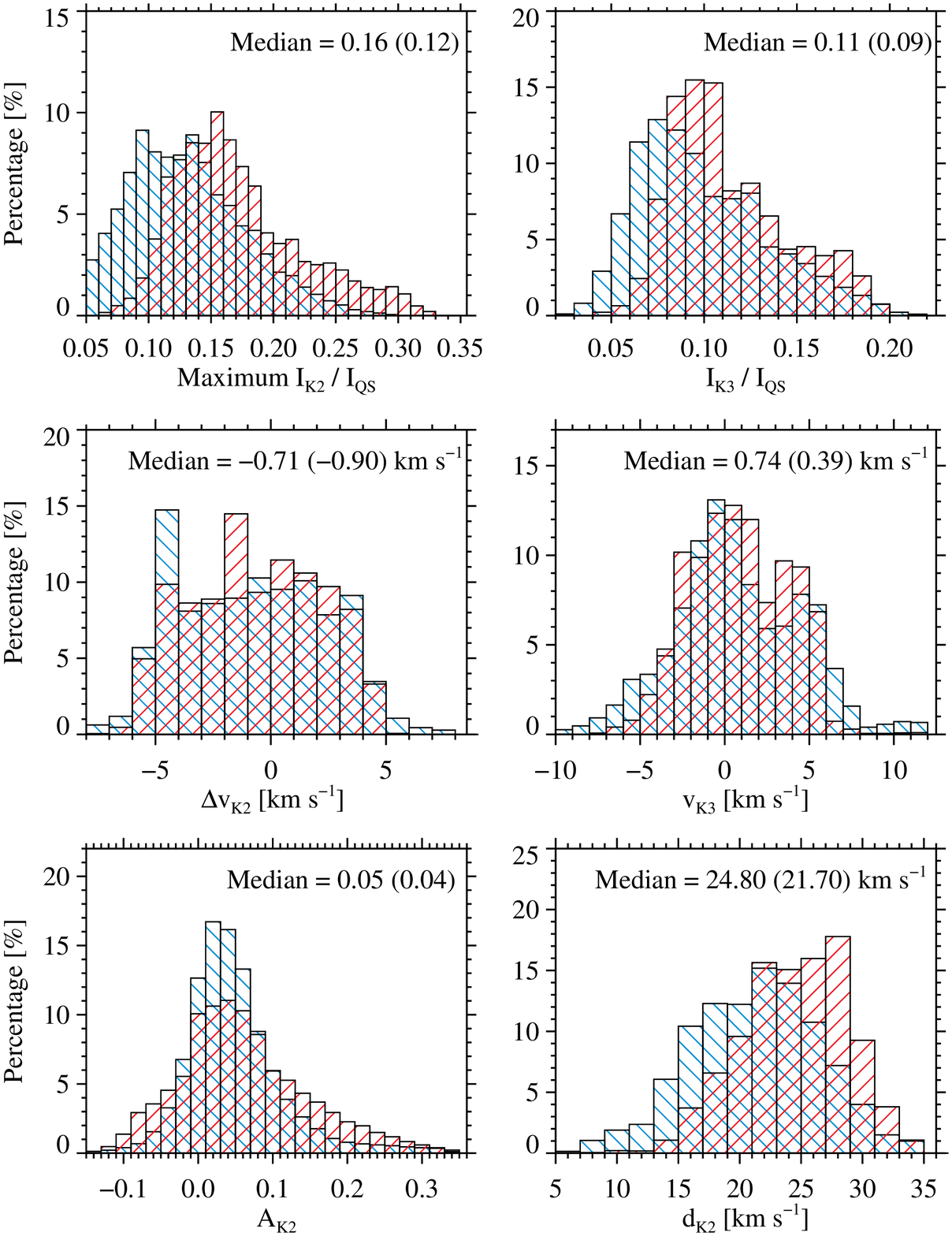} 
\caption{Histograms of maximum K$_{2}$~peak intensity, K$_{3}$~peak intensity, average Doppler shift of the K$_{2}$~peaks, Doppler shift of the K$_{3}$~peak, asymmetry of the K$_{2}$~peaks and their separation. Red distributions stand for pixels within PMJs, blue for their 3\arcsec $\times$ 3\arcsec\, background. The median values of the red (blue) distributions are indicated in the upper corner of the panels.}
\label{fig:histo_cak_pmj}
\end{figure}

Finally, we have performed a statistical analysis of the parameters inferred from the \ion{Ca}{2}~K line to characterize some properties of the PMJ in the chromosphere. This statistical analysis is based on 14300~pixels belonging to the detected PMJ features. We only considered the time step corresponding to the maximum brightness of each PMJ feature during its temporal evolution. Although the individual evolution of each pixel may slightly differ, most of the pixels within the PMJ are at their maximum brightness. In addition, we compared the results with the properties of the surroundings. The surroundings was defined as those pixels in the 3\arcsec $\times$ 3\arcsec \, area centered on the PMJ that do not belong to the PMJ. This second set consists of 171500 pixels. Unsuccessful identifications of the profile features were discarded from both sets.
\begin{table}[t]
\caption{Spectral lines and nodes in temperature, LOS velocity, microturbulence, longitudinal and horizontal components and azimuth of the magnetic field used in each cycle of inversions. Cycles with '$\dagger$' symbols were performed over a downsampled FOV. Star superscripts denote non-equidistant nodes, specifically located at log($\tau$) [--7.6, --5.0, --3.8, --2.6, --1.5, 0.5], [--7.6, --5.0, --3.8, --2.8, --1.9, --1.2, 0.5] and [--5.8, --5.0, --3.8, --2.8, --1.9, --1.2, 0.5] in LOS velocity, [--5.5, --2.5, --0.5] in B$_{||}$, and [--5.5, --0.5] in B$_{\perp}$ and $\phi$.}
\begin{tabular}{lccccc}
\textbf{cycle} & \textbf{Spectral lines} & \textbf{T} & \textbf{v$_{LOS}$}, \textbf{v$_{mic}$} & \textbf{B$_{||}$, B$_{\perp}$, $\phi$} \\ 
\hline
\hline
1$^{\dagger}$ & \ion{Fe}{1}, \ion{Ca}{2}~IR & 7 & 5, 3 & 2, 1, 1 \\ 
2 & \ion{Fe}{1}, \ion{Ca}{2}~IR & 7 & 6, 3 & 2, 1, 1 \\ 
3$^{\dagger}$ & \ion{Fe}{1}, \ion{Ca}{2}~IR, \ion{Ca}{2}~K & 7 & 6, 3 & 2, 1, 1 \\  
4 & \ion{Fe}{1}, \ion{Ca}{2}~IR, \ion{Ca}{2}~K & 7 & 6$^{\star}$, 3 & 3, 2, 1 \\ 
5 & \ion{Fe}{1}, \ion{Ca}{2}~IR, \ion{Ca}{2}~K & 7 & 6$^{\star}$, 4 & 3, 2, 2 \\ 
6 & \ion{Fe}{1}, \ion{Ca}{2}~IR, \ion{Ca}{2}~K & 7 & 7$^{\star}$, 4 & 3$^{\star}$, 2$^{\star}$, 2$^{\star}$ \\ 
\hline \\
\end{tabular}
\label{tab:cycles}
\end{table}

Figure~\ref{fig:histo_cak_pmj} shows histograms of the maximum K$_{2}$~peak intensity, the K$_{3}$~peak intensity, and the velocity diagnostics given by the \ion{Ca}{2}~K line for both sets, which are normalized to the number of pixels in each sample. The distributions of the maximum K$_{2}$~and K$_{3}$~peak intensities diverge between groups. Both sets show wide distributions but pixels within PMJs have greater values, mainly in the maximum K$_{2}$~peak intensity that reaches up to 0.32. The median maximum K$_{2}$~and K$_{3}$~peak intensities in PMJs are 0.16 and 0.11, respectively. Most of the pixels of both sets show average Doppler shifts of the K$_{2}$~peaks and Doppler shifts of the K$_{3}$~peak less than $\pm$4~km~s$^{-1}$. However, the latter quantity tends to be more redshifted in comparison to the former. PMJs harbor median average Doppler shifts of the K$_{2}$ peaks and Doppler shifts of the K$_{3}$~peak of --0.71 and 0.74~km~s$^{-1}$. Although the distributions of the asymmetry of the K$_{2}$~peaks in PMJs and their surroundings have similar shapes, the former one is wider, where most values are around --0.08 and 0.25, and its median value is 0.05. Both distributions are shifted to positive asymmetries, indicating that flows tend to harbor more redshifted velocities between the K$_{2}$ and K$_{3}$ peak formation heights. Regarding the separation of the K$_{2}$~peaks, PMJs usually show greater values than their surroundings and their median value is 24.8~km~s$^{-1}$.

Therefore, PMJs are observed as brighter locations in the K$_{2}$ and K$_{3}$ peaks that usually stand out in the K$_{2}$~peak separation.

\begin{figure}[!t]
\centering
\includegraphics[trim = {2.8cm 4.2cm 2.8cm 3cm}, clip, width = 0.41\textwidth]{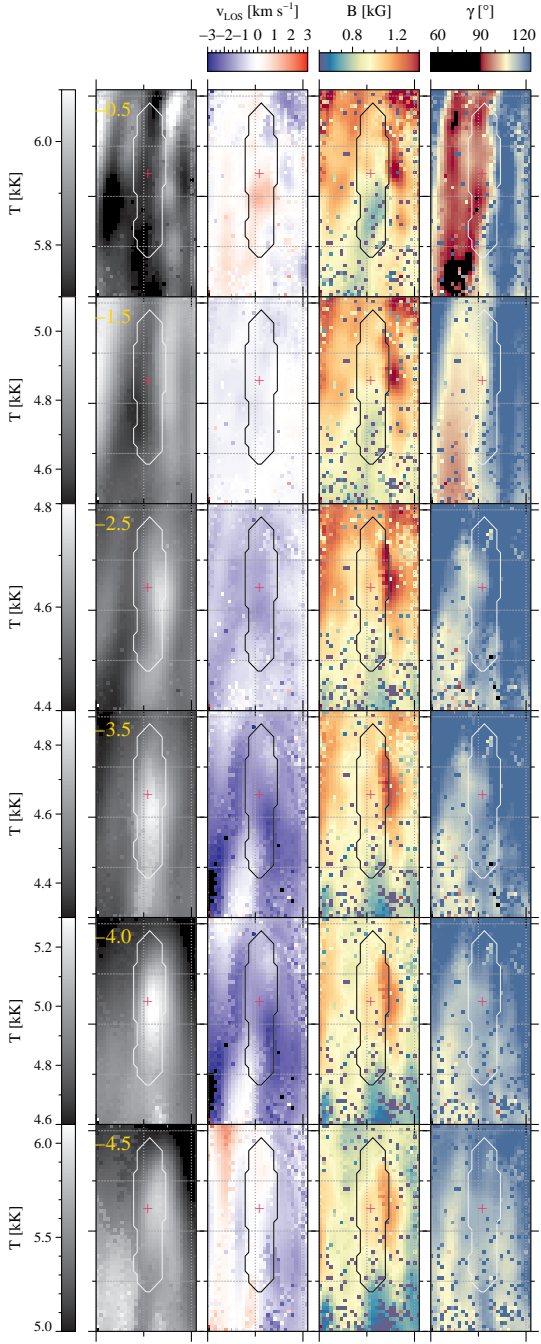} 
\caption{Temperature, LOS velocity, magnetic field strength and inclination obtained from inversions of PMJ~1, which is enclosed by contours. Height from solar surface increases from top to bottom, as indicated by the log($\tau$) values in the temperature maps. Major tick marks represent 1\arcsec. The plus symbol marks the pixel represented in Figure~\ref{fig:ajuste_stic_pmj1}.}
\label{fig:inv_pmj1}
\end{figure}

\subsection{Inversions of PMJs}

We performed an inversion of the Stokes profiles in the \ion{Fe}{1}~630~nm pair, \ion{Ca}{2}~854.2~nm and \ion{Ca}{2}~K lines emerging from PMJs with the STiC code \citepads{2016ApJ...830L..30D, 2018arXiv181008441D} to retrieve atmospheric information within the formation regions of these spectral lines. Although it would be interesting to analyze all detected PMJs, we considered the PMJs~1 and~2 since these inversions are computationally expensive.

STiC is a new non-LTE inversion code based on the RH code \citepads{2001ApJ...557..389U} to solve the non-LTE problem. For a given stratified atmospheric input model, it derives pressure scales considering hydrostatic equilibrium. In addition, it assumes a plane-parallel geometry to calculate the atomic population in each pixel and considers the effect of partial redistribution in angle and frequency of scattered photons \citepads{2012A&A...543A.109L}. The radiative transport equation is solved using cubic Bezier solvers \citepads{2013ApJ...764...33D}. The inversion engine of STiC includes an equation state extracted from the SME code \citepads{2017A&A...597A..16P}. 

The same strategy was followed in both examples that consists of six cycles of inversions, as is summarized in Table~\ref{tab:cycles}. We initialized our inversions employing a simple input model, whose physical parameters are commonly found in sunspots penumbrae (LOS velocity of 1~km~s$^{-1}$, magnetic field strength of 1~kG and inclination of 130$\degree$) and its temperature imitates the height dependence between the photosphere and upper chromosphere \citepads[as in]{1993ApJ...406..319F}, over a downsampled FOV enclosing the PMJ and its surroundings. The use of a downsampled FOV is supported by the smooth spatial distribution of the profiles in PMJs (Section~\ref{sub:spatialdistribution}), which suggests that physical conditions do not abruptly vary within them. We use the resized version of the output model as input model in the following cycle.

\begin{figure}[!t]
\centering
\includegraphics[trim = {2.5cm 0.15cm 10cm 2.8cm}, clip, width = 0.48\textwidth]{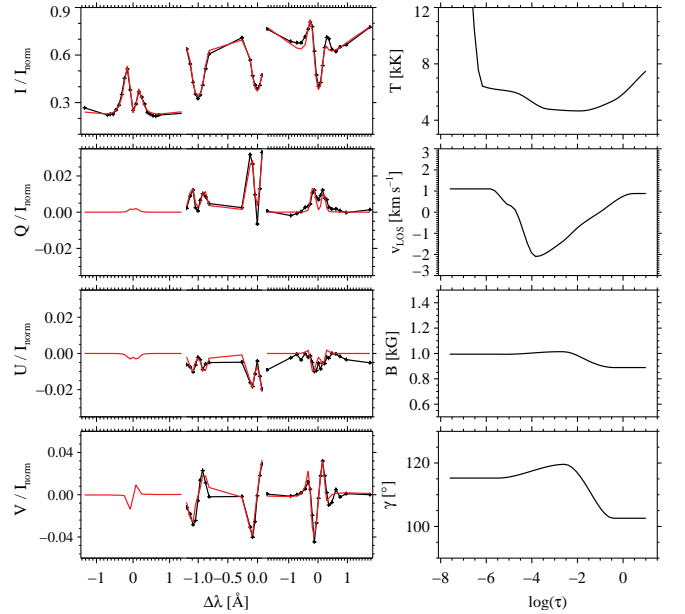} 
\caption{Results inferred from a pixel within PMJ 1 (plus symbol in Figure~\ref{fig:inv_pmj1}). \textit{Left}: Observed and best-fit profiles (black and red lines). The x-axes represent the calibrated wavelengths, which are sorted into ascendent order: \ion{Ca}{2}~K, the \ion{Fe}{1}~630~nm pair and \ion{Ca}{2}~854.2~nm. The y-axes are given in CGS units (erg~cm$^{-2}$~s$^{-1}$~ster$^{-1}$). The Stokes~$I$ profile of \ion{Ca}{2}~K is multiplied by 3 for better visibility. \textit{Right}: Atmospheric parameters resulting from the inversion of the Stokes profiles shown on the left side.}
\label{fig:ajuste_stic_pmj1}
\end{figure}

To speed up convergence, the first cycles were performed without \ion{Ca}{2}~K data. Furthermore, we used a slightly smoothed version of the output model obtained in the previous cycle as the input model of the next one when consecutive cycles have the same FOV. 

\subsubsection{PMJ 1}

Figure~\ref{fig:inv_pmj1} shows the atmosphere retrieved from inversions of PMJ~1 at different heights from the photosphere to the mid chromosphere. The achieved fits are good in most of the profiles, as Figure~\ref{fig:ajuste_stic_pmj1} shows for a pixel inside the PMJ where the fits capture the shape of the observed Stokes profiles within the noise. The inferred atmospheric parameters are also displayed. Although the plots extend to log($\tau$)=--7.6, we only consider results obtained up to heights that are equivalent to log($\tau$)=--4.5 since the uncertainties in the inverted parameters are large for greater heights. 

The temperature at log($\tau$)=--0.5 displays typical bright penumbral filaments embedded in a dark photospheric background, with these features appearing more blurred at log($\tau$)=--1.5. However, the temperature at the center of the PMJ increases considerably from 4700 to 6000~K at higher layers, which is about 200--500~K greater as compared to the surroundings.

The second column of Figure~\ref{fig:inv_pmj1} shows photospheric flow channels harboring LOS velocities similar to those inferred from the ME inversions, even a roundish downflow appears at approximately the same place as in Figure~\ref{fig:photo_pmjs}. In the upper layers, LOS velocities are predominantly blueshifted and enhanced between log($\tau$)=--2.5 and --4, though we can distinguish small variations. A strong blueshift is visible above the photospheric downflow until log($\tau$)=--4, but not further up. A less blueshifted region, of about --1~km~s$^{-1}$, partially overlaps the inner end of the PMJ at --3.5 and becomes more redshifted with height until log($\tau$)=--4.5, where most of the PMJ is at rest. Despite LOS velocities at log($\tau$)=--4.5 differ from those computed from the K$_3$~peak, the general qualitative behavior agrees with that inferred from the \ion{Ca}{2}~K diagnostics. 

Lastly, the upper row of the two last columns of Figure~\ref{fig:inv_pmj1} shows a spine-intraspine region, as in Figure~\ref{fig:photo_pmjs}. The magnetic field strength and inclination are weaker and more diffuse at higher layers. However, besides the strong magnetic field on the right side, magnetic fields of $\sim$1~kG are visible within the PMJ at log($\tau$)=--2.5, which rapidly weaken to 0.6~kG along it, as the corresponding polarization signals do (Figure~\ref{fig:example1}). These values are reduced in the chromosphere, albeit remaining discernible. The field lines within the PMJ are quite horizontal ($\sim$115$\degree$), suggesting that it follows the field lines of the chromospheric canopy. Therefore, the perpendicular component of the magnetic field contributes significantly to the field strength and explains the difference between the results derived from the inversions and the WFA (see Section~\ref{sub:pol}).

\begin{figure*}[!t]
\centering
\includegraphics[trim = {2.4cm 2.9cm 2.8cm 2.2cm}, clip, width = 0.96\textwidth]{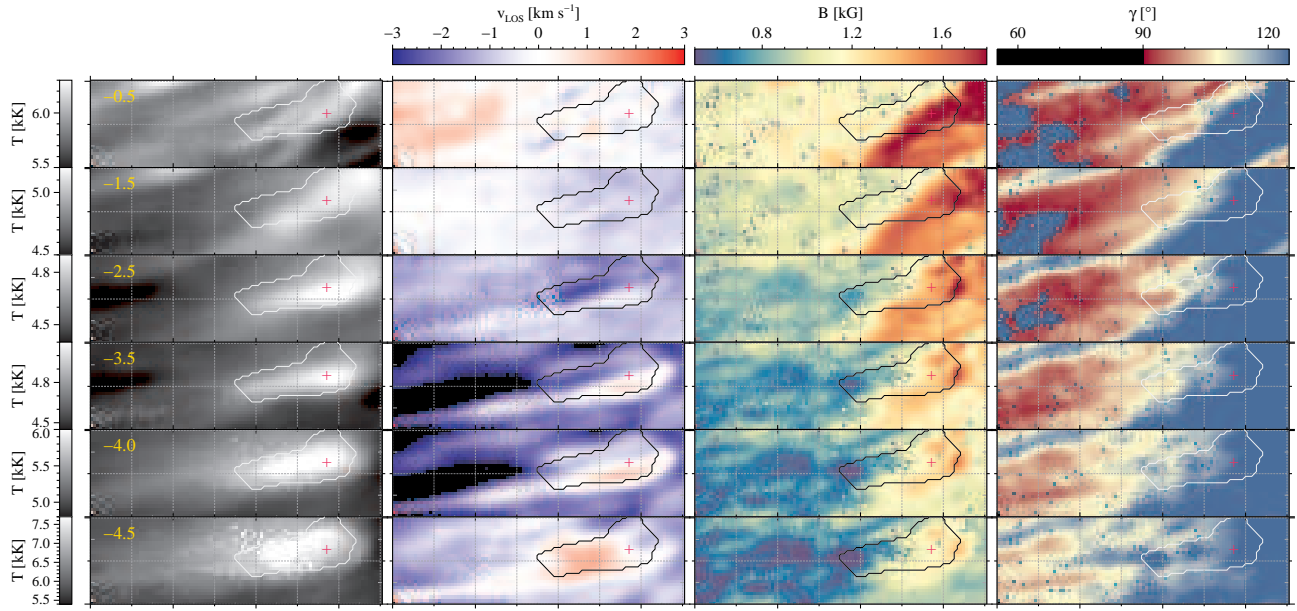} 
\caption{Temperature, LOS velocity, magnetic field strength and inclination inferred from inversions of PMJ~2. The layout is the same as in Figure~\ref{fig:inv_pmj1}. The plus symbol marks the pixel whose results are shown in Figure~\ref{fig:ajuste_stic_pmj4}.}
\label{fig:inv_pmj4}
\end{figure*}

\begin{figure}[!t]
\centering
\includegraphics[trim = {2.6cm 0.6cm 10cm 3.3cm}, clip, width = 0.48\textwidth]{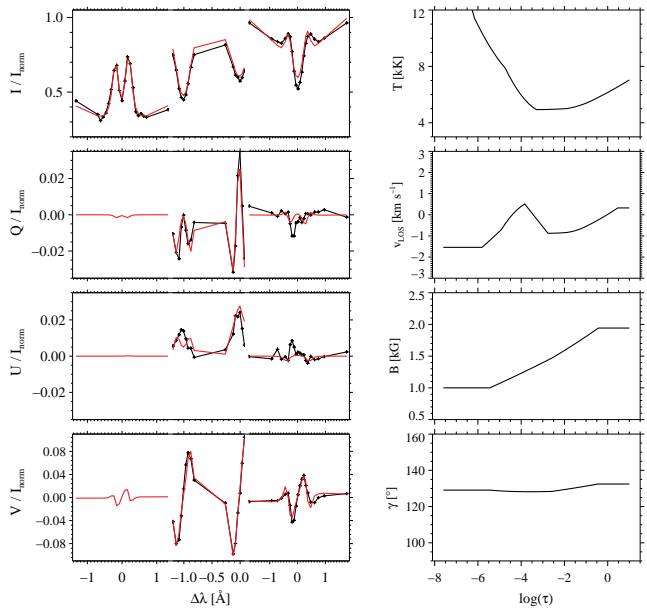} 
\caption{Results obtained from a pixel within PMJ~2 (plus symbol in Figure~\ref{fig:inv_pmj4}). The layout is the same as in Figure~\ref{fig:ajuste_stic_pmj1}.}
\label{fig:ajuste_stic_pmj4}
\end{figure}

\subsubsection{PMJ 2}

The atmospheric parameters inferred from PMJ~2 are presented in Figure~\ref{fig:inv_pmj4}. As in the case of PMJ~1, the inversions of most of the pixels were sucessful, as the fits achieved for the observed Stokes profiles illustrate in Figure~\ref{fig:ajuste_stic_pmj4}. 

Generally, the physical parameters shown in Figure~\ref{fig:inv_pmj4} reveal variations akin to those depicted in PMJ~1. The temperature enhances within the PMJ at log($\tau$) $=$ --2.5 and above, reaching up to $\sim$7500~K at log($\tau$)=--4.5. Regarding the LOS velocity, the hot feature is partially covered by an almost at rest region embedded in a mainly blueshifted background at log($\tau$)=--2.5. This region is more redshifted with height, reaching up to 2 km~s$^{-1}$ at log($\tau$)=--4.0. A strong and elongated upflow is visible on the left side of the FOV, which likely belongs to the fibril partially covering this PMJ (Figure~\ref{fig:example4time}). LOS velocities do not show any relation with temperature. At log($\tau$)=--4.5, the FOV is more redshifted and a roundish upflow pops up at the inner end. This agrees with the \ion{Ca}{2}~K line diagnostics (Figure~\ref{fig:obscak_pmj4}) though LOS velocities are usually more positive than those computed from the K$_{3}$~peak, probably due to sensitivity extrapolations during the inversions although we tried different node configurations to avoid them.

As before, this PMJ is co-spatial with a spine-intraspine boundary in the photosphere that is more diffuse with increasing height. A magnetic structure coinciding with the PMJ is discernible at log($\tau$)=--3.5, which weakens with height. This PMJ also reveals a magnetic field strength gradient along it, with strengths of 1.3~kG in the inner end that lessen until 0.6~kG in the outer one, and it is related to horizontal field lines of about 110$\degree$ inclination.

\section{Discussion}
\label{sec:discussion}

Our study provides information on the polarization signals and the physical parameters of PMJs detected in high spatial resolution data. We also revisit some aspects previously reported, such as their morphology and appearance. In this section we discuss our results to highlight how they broaden our knowledge on PMJs. 

\subsection{Polarization and physical parameters of PMJs}

High spatial resolution data reveal PMJs as well-extended and spatially resolved features harboring clear intensity and polarization signals in the chromosphere. 

All the detected PMJs are located above photospheric regions where magnetic field inclinations have significative horizontal gradients, such as at the spine-intraspine boundaries or the outer edges of penumbral filaments and the sunspot outer boundary. These regions are susceptible of harboring field lines that are pulled down by (convective) downflows, as several observational studies have found \citepads{2011LRSP....8....4B, 2013A&A...553A..63S, 2013A&A...549L...4R, 2016A&A...596A...4F}. As \citetads{2016ApJ...816...92T} proposed, magnetic reconnection is likely to occur in an environment with large magnetic field variations taking place at such small spatial scales. Reconnection products could be driven to upper layers through a region of reduced magnetic field formed between the spot field lines dragging to the solar surface and those remaining steady. Even though we do not detect clear evidence of magnetic reconnection, its existence can not be ruled out as they could happen on scales smaller than our spatial resolution or in deeper atmospheric layers.

The apparent rise speeds reported so far are of order 100~km~s$^{-1}$ \citepads{2007Sci...318.1594K, 2015ApJ...811L..33V, 2016ApJ...816...92T}. Assuming that these speeds are associated with vertical motions of gas, the Doppler velocities inferred from PMJs should be of the same order of magnitude, as the heliocentric angle is small. On the other hand, if they were related to horizontal motions, PMJs would systematically harbor LOS velocities of about 15~km~s$^{-1}$. However, our results reveal that these features do not stand out from their surroundings because of their LOS velocities, as most of them harbor values of less than $\pm$4~km~s~$^{-1}$, which is consistent with the absence of strong lineshifts in their profiles.

Considering the appearance of the emission peaks of PMJ-like \ion{Ca}{2}~854.2~nm intensity profiles, and also that of the K$_{2}$~peaks in most of the cases, we identify five types of intensity profiles along PMJs that correspond to different velocity gradients along their LOS irrespective of their locations within the spot. Thence, we can not allocate a unique vertical velocity in the chromosphere to the PMJs. 

In any case, we find LOS velocities that are much lower than the apparent rise speeds of PMJs reported in the literature (typically in excess of 100~km~s$^{-1}$). This discrepancy between measured LOS velocities and apparent motions indicates that the fast appearance of PMJs cannot be explained by plasma flows alone, similarly to type II spicules. At the limb, type II spicules display apparent rise velocities of  30--150~km~s$^{-1}$ \citepads{2007PASJ...59S.655D, 2012ApJ...759...18P} while their TR counterparts display apparent speeds on the disk of 80--300~km~s$^{-1}$ \citepads{2014Sci...346A.315T, 2016SoPh..291.1129N}. Doppler shift measurements however, are typically in the range 20--50~km~s$^{-1}$ \citepads{2009ApJ...705..272R, 2012ApJ...752..108S} or 50--70~km~s$^{-1}$ in TR diagnostics \citepads{2015ApJ...799L...3R}.

Recently, \citetads{2017ApJ...849L...7D} explain the discrepancy observed in type II spicules by the fast apparent motion being caused by a heating front moving at much higher velocity than the actual mass flow. Magnetic concentrations of different strength and inclination, due to convective motions in the photosphere below the spicule, build up currents that penetrate into the chromosphere within the spicule, where they are rapidly dissipated at Alfv\`{e}nic speeds by ambipolar diffusion, producing in turn rapid heating. Therefore, the apparent high-speed motion in type II spicules is due to both effects: fast propagation of currents and rapid associated heating.

We speculate that a similar scenario might be at play in PMJ formation. Magnetic reconnection in the deep photosphere of the penumbra may trigger fast currents that act as an upward propagating perturbation front. Their propagation might produce current heating and, therefore, increase the temperature higher in the atmosphere. The distinctive emission peaks in the PMJ \ion{Ca}{2}~854.2~nm intensity profiles and the enhanced K$_{2}$ peak separation found in PMJs suggest temperature enhancements taking place at the temperature minimum region and low chromosphere. This heating may continue through the chromosphere as the atmospheric model retrieved from the STiC inversions reveals, where PMJs are related to hot features that expand and are hotter at larger height, reaching up to 7500~K at log($\tau$) = -4.5. Furthermore, the overall enhanced intensity in the PMJ \ion{Ca}{2}~K profiles may be explained by a relatively increased coupling to the local conditions in the higher atmosphere as a result from enhanced collisions with electrons associated to the currents.

In addition, PMJs exhibit enhanced polarization signals in \ion{Ca}{2}~854.2~nm, specifically in circular polarization. Our inversions indicate that this polarization increase in PMJs might be related to magnetic fields of about 1~kG in the chromosphere. Considering the inferred average magnetic field strengths and densities within the PMJ at log($\tau$) = -2.5, -3.0, -3.5 and -4.0, we retrieve Alfv\'{e}n speeds of 25, 35, 55 and 150 km~s$^{-1}$ at those optical depths. Therefore, the apparent speeds reported on before are approximately of the same order of magnitude. Finally, results suggest that field lines in PMJs are more horizontal in the upper layers, in agreement with the inclination height variation found by \citetads{2008A&A...488L..33J}.

\subsection{Comparison with previous studies}

Statistics on the spatial dimensions of the detected PMJs give us values akin to those found by \citetads{2007Sci...318.1594K}, but greater than those reported by \citetads{2017A&A...602A..80D}. The latter could be due to differences in the detection method, as we have probably missed smaller PMJs. Nevertheless, our results are comparable. Regarding their duration, PMJs have lifetimes similar to those measured in \citetads{2017A&A...602A..80D}. We also find hotspots as in \citetads{2016ApJ...816...92T} and \citetads{2017A&A...602A..80D}, where PMJs recursively appear.

Usually, PMJs are also visible in the mid-upper chromosphere since they appear as brightenings or small bulges inside bright fibrils in the \ion{Ca}{2}~854.2~nm and \ion{Ca}{2}~K line cores filtergrams, in agreement with the findings of \citetads{2015ApJ...811L..33V}. 

The most common types of \ion{Ca}{2}~854.2~nm intensity profiles are those harboring a blue emission peak, which emerge alone or together with the red one, as reported by \citetads{2017A&A...602A..80D}. PMJs are usually affected by other neighbor features, such as short dynamic fibrils, that cover up their emission signals.

In the photosphere, observables computed from the \ion{Fe}{1}~630.15~nm line, such as Dopplergrams, NCP maps and red-wing magnetograms, exhibit diverse signals below the detected PMJs. However, they seem to be related to changes in the photospheric penumbra as their apparent inclination is similar to that of the penumbral filaments, regardless the PMJ evolution. In addition, these patches are not exclusively located within the transients. Thus, as opposed to \citetads{2010A&A...524A..20K} and \citetads{2010A&A...524A..21J}, we can not unambiguously relate these signals to PMJs. To check if these transients leave any imprint in the photosphere, a more detailed analysis of the photospheric data is needed.

\section{Summary and conclusions}
\label{sec:summary}

For the first time, we have characterized the polarization signals emerging from PMJs and have inferred physical parameters in their atmospheres. This has been achieved thanks to the CHROMIS and CRISP instruments that allow us to acquire photospheric and chromospheric information through high spatial, spectral and temporal resolution data. Considering observed properties reported on earlier, we have manually identified 37~PMJs on both penumbral sides at different radii. Particularly, many of them emerge in the outer penumbral boundary, and even outside the spot in rare ocassions. 

PMJs are well-defined features that appear above locations where the photospheric magnetic field inclination shows a sheared configuration, such as at the boundary between spines and intraspines. This leads us to conclude that they require strong field inclination gradients to occur. There, some field lines are submerged allowing their reconnection with spines in deeper layers. Reconnection products might be led to upper layers through reduced magnetic field conduits that appear due to the bending of the photospheric field lines. This scenario is similar to that proposed by \citetads{2016ApJ...816...92T}. However, magnetic reconnections might take place below the photosphere or on scales smaller than our spatial resolution and, consequently, their imprints may have eluded direct detection. 

By performing inversions with the STiC code, or by other analysis of the \ion{Ca}{2}~K line profiles, PMJs are concluded to be associated with small LOS velocities compared with the high apparent rise speeds found in previous studies \citepads{2007Sci...318.1594K, 2015ApJ...811L..33V, 2016ApJ...816...92T}. Furthermore, it is not univocal to designate an exclusive behavior of their vertical velocity as diverse gradients can be inferred depending on the asymmetries of the emission peaks of the \ion{Ca}{2}~854.2~nm intensity profiles and that of the K$_{2}$~peaks. This suggests that PMJs might not be primarily related to actual gas motions induced by magnetic reconnections. 

Similarly to the explanation for the type II spicules given by \citetads{2017ApJ...849L...7D}, PMJs could be due to the propagation of perturbation fronts originated by magnetic reconnections in the deep photosphere. A plausible possibility is that these perturbations are currents that dissipate energy within the PMJ. Having more electrons can also increase collisional rates locally and, thus, enhance the coupling to the local conditions during this transient phase. This can also contribute to explaining the overall brighter PMJ \ion{Ca}{2}~K profiles.

The heating might continue until, at least, TR temperatures, as \citetads{2015ApJ...811L..33V} found. Our results support this idea as the greater K$_{2}$~peak separations found in PMJs indicate temperature rises in the low chromosphere and, moreover, inversions reveal that PMJs expand and heat up with height, from 4500~K to about 7500~K.

Polarization signals within PMJs stand out from their surroundings in the \ion{Ca}{2}~854.2~nm line, specifically in circular polarization that usually shows multi-lobed profiles mostly due to the peculiar shape of Stokes~$I$. Inversions results show that PMJs are related to magnetic features of $\sim$1~kG that are more horizontal at larger heights, in agreement with \citetads{2008A&A...488L..33J}.

Finally, most of this analysis has been carried out by taking into account only the moment when PMJs are at their maximum brightness. A natural step to continue this work would be to study the temporal evolution of these transients in order to shed more light on them.

\medskip 
\small \textit{Acknowledgements.} We are grateful to B. De Pontieu and J. Leenaarts for helpful comments. This study has been discussed in the workshop '\textit{Studying magnetic-field-regulated heating in the solar chromosphere}' (team 399) at the International Space Science Institute (ISSI) in Switzerland. This research has been supported by the Knut and Alice Wallenberg Foundation, the Research Council of Norway (project number 250810) and through its Centres of Excellence scheme (project number 262622). JdlCR is supported by grants from the Swedish Research Council (2015-03994), the Swedish National Space Board (128/15) and the Swedish Civil Contingencies Agency (MSB). This project has received funding from the European Research Council (ERC) under the European Union's Horizon 2020 research and innovation programme (SUNMAG, grant agreement 759548). This paper is based on data acquired at the Swedish 1-m Solar Telescope, operated by the Institute for Solar Physics of Stockholm University in the Spanish Observatorio del Roque de los Muchachos of the Instituto de Astrof\'{\i}sica de Canarias. The Institute for Solar Physics is supported by a grant for research infrastructures of national importance from the Swedish Research Council (registration number 2017-00625). Our computations were performed on resources provided by the Swedish National Infrastructure for Computing (SNIC) at the High Performance Computing Center North at Ume\aa\, University for Kebnekaise with project id SNIC2017-11-17. This research has made use of NASA's Astrophysical Data System.

\bibliographystyle{apj}
\bibliography{biblioads}

\end{document}